\documentclass[sigconf]{acmart}
\PassOptionsToPackage{bookmarks={false}}{hyperref}
\AtBeginDocument{%
  \providecommand\BibTeX{{%
    \normalfont B\kern-0.5em{\scshape i\kern-0.25em b}\kern-0.8em\TeX}}}

\hypersetup{bookmarksdepth=-2}
\usepackage{url}            % simple URL typesetting
\usepackage{booktabs}       % professional-quality tables
\usepackage{amsfonts}       % blackboard math symbols
\usepackage{nicefrac}       % compact symbols for 1/2, etc.
\usepackage{microtype}      % microtypography
\usepackage{mathrsfs}
\usepackage{textcomp, graphicx, subfigure}
\usepackage{epsfig}
\usepackage{amssymb}
\usepackage{amsmath}
\usepackage{algorithmic}
\usepackage{algorithm}
\usepackage{multirow}
\usepackage{ulem}
\usepackage{amssymb}
\usepackage{mathtools,bbm}
\usepackage{enumitem}
\usepackage{color,soul}
\usepackage{xcolor}
\usepackage{multicol}

\usepackage{natbib}
\usepackage{adjustbox}

\copyrightyear{2020}
\acmYear{2020}
\setcopyright{acmcopyright}\acmConference[ICCAD '20]{IEEE/ACM International Conference on Computer-Aided Design}{November 2--5, 2020}{Virtual Event, USA}
\acmBooktitle{IEEE/ACM International Conference on Computer-Aided Design (ICCAD '20), November 2--5, 2020, Virtual Event, USA}
\acmPrice{15.00}
\acmDOI{10.1145/3400302.3415629}
\acmISBN{978-1-4503-8026-3/20/11}

\begin{document}
\title{SF-GRASS: \emph{S}olver-\emph{F}ree {\emph{Gra}ph} \emph{S}pectral \emph{S}parsification}

\author{Ying Zhang }
\authornote{Equal contribution}
\affiliation{
% \institution{Department of ECE}
\institution{Stevens Institute of Technology}
\city{Hoboken}
\state{New Jersey}
\postcode{07030}
}
\email{yzhan232@stevens.edu}

\author{Zhiqiang Zhao }
\authornotemark[1]
\affiliation{
% \institution{Department of ECE}
\institution{Michigan Technological University}
\city{Houghton}
\state{Michigan}
\postcode{49931}
}
\email{qzzhao@mtu.edu}

\author{Zhuo Feng}
\affiliation{
% \institution{Department of ECE}
\institution{Stevens Institute of Technology}
\city{Hoboken}
\state{New Jersey}
\postcode{07030}
}
\email{zhuo.feng@stevens.edu}
\begin{abstract}
Recent spectral graph sparsification techniques have shown promising performance in accelerating many numerical and graph algorithms, such as iterative methods for solving large sparse matrices, spectral partitioning of undirected graphs, vectorless verification of power/thermal grids, representation learning of large graphs, etc. However, prior spectral graph sparsification methods rely on fast Laplacian matrix solvers that are usually challenging to implement in practice. This work, for the first time, introduces a solver-free approach (SF-GRASS) for spectral graph sparsification by leveraging emerging spectral graph coarsening and  graph signal processing (GSP) techniques. We introduce a  local spectral embedding scheme for efficiently identifying spectrally-critical edges that are key to preserving graph spectral properties, such as the first few Laplacian eigenvalues and eigenvectors. Since the   key kernel functions in SF-GRASS can be efficiently implemented using sparse-matrix-vector-multiplications (SpMVs), the proposed spectral approach is simple to implement and inherently parallel friendly. Our extensive experimental results show that the proposed method can produce a hierarchy of high-quality spectral   sparsifiers in nearly-linear time  for a variety of real-world, large-scale graphs and circuit networks when compared with prior state-of-the-art spectral methods.
\end{abstract}






\maketitle

% \vspace{-0.3cm}

% \keywords{Spectral graph theory, graph partitioning, sparse matrix solver}
%  \vspace{-0.3cm}

\section{Introduction}
{Spectral methods are playing increasingly important roles in a wide variety of graph and numerical applications}~\cite{teng2016scalable}. Examples include scientific computing and numerical optimization~\cite{spielman2014sdd,kelner2014almost,feng2016spectral}, graph partitioning and data clustering~\cite{lee2014multiway, peng2015partitioning}, machine learning and data mining~\cite{kipf2016semi,deng2019graphzoom}, as well as integrated circuit modeling, simulation and verifications \cite{xueqian:tcad15,lengfei:tcad15,zhiqiang:dac17}. 
In particular, latest theoretical breakthroughs in spectral graph theory have led to the development of nearly-linear time spectral graph sparsification \cite{spielman2011spectral,feng2016spectral,Lee:2017,zhuo:dac18} and coarsening algorithms \cite{loukas2018spectrally,loukas2019graph,zhao2018nearly, zhao:dac19}. These techniques can efficiently produce much smaller graphs that well preserve the key spectral properties of the original graph (e.g., the first few eigenvalues and eigenvectors of the graph Laplacian), which in turn has led to much faster algorithms for solving partial differential equations (PDEs) and linear systems of equations \cite{spielman2011spectral,peng2013phd,zhiqiang:iccad17}, spectral clustering and graph partitioning \cite{peng2015partitioning,lee2014multiway,zhuo:dac18,zhao:dac19}, and dimensionality reduction and data visualization \cite{zhao2018nearly}. 

However, prior spectral graph sparsification methods   strongly rely on fast Laplacian matrix solvers that are usually challenging to implement in practice and inherently-difficult accelerate on parallel processors. For example, effective-resistance sampling-based spectral sparsification method \cite{spielman2011graph} requires multiple Laplacian matrix solutions for computing each edge's leverage score, while the latest spectral-perturbation based algorithm \cite{feng2020grass} leverages a graph-theoretic algebraic multigrid (AMG) solver for computing dominant generalized eigenvectors key to estimating each edge's spectral importance. As a result, the performance (scalability) of   Laplacian matrix solver can become a dominating factor in existing spectral sparsification methods. However, after  decades of extensive research studies by theoretical computer scientists, it is still not   clear if there exist any practically-efficient (nearly-linear time) and robust  Laplacian solvers for general large-scale real-world graphs.
 
This paper for the first time introduces a solver-free spectral graph sparsification framework (SF-GRASS) by leveraging emerging spectral graph coarsening \cite{zhao:dac19} and graph signal processing techniques \cite{shuman2013emerging}. Our approach first coarsens the original graph into increasingly smaller graphs while preserving the key graph spectral properties.  Since  spectral  graph  coarsening   \cite{zhao:dac19} can be considered as a cascade of low-pass graph filters with decreasing bandwidths, spectrally-critical edges for different ranges of eigenvalues can be effectively identified on coarse-level graphs in a stratified manner. For example,   considering a coarsest graph that has only a few (e.g. two) nodes, any  graph signals smoothed (low-pass filtered) from random vectors can become good approximations of the Fiedler vector corresponding to the few smallest nontrivial Laplacian eigenvalues; when such vectors are leveraged for recovering spectrally-critical edges  using spectral-perturbation based approach similar to the one introduced in \cite{feng2016spectral}, ultra-sparse spectral graph sparsifiers preserving the smallest few eigenvalues can be efficiently extracted; by iteratively mapping   sparsifiers back to each finer level, a hierarchy of spectral sparsifiers with increasing sizes can be  incrementally computed  for preserving increasing eigenvalues. Our results show SF-GRASS outperforms prior state-of-the-art methods for spectral sparsification considering both efficiency and solution quality. The  technical contribution of this work has been summarized as follows:
\begin{enumerate}
  \item For the first time, we present a solver-free  spectral graph sparsification framework  (SF-GRASS)  by leveraging emerging spectral graph coarsening \cite{zhao:dac19} and graph signal processing techniques \cite{shuman2013emerging}. It can be implemented using simple sparse-matrix-vector multiplications and thus completely addresses the computational challenges in prior methods that strongly reply on  efficient graph Laplacian solvers. 
    \item We introduce a multilevel spectral sparsification framework, which is motivated by the prior graph spectral perturbation  analysis approach \cite{feng2016spectral}. Such a scalable framework allows constructing a hierarchy of spectrally-reduced and sparsified graphs in nearly-linear time, which can become key to accelerating many graph-based numerical computing tasks.
    \item  By comprehensively comparing with the state-of-the-art method through extensive experiments, we show that  in various numerical and graph-related applications, such as solving sparse SDD matrices, and vectorless verification of power grids, SF-GRASS can always obtain high-quality solution while achieving dramatically improved runtime scalability.  
\end{enumerate}
The rest of this paper is organized as follows.  Section
\ref{background_sec} provides a brief introduction to   spectral graph sparsification and coarsening problems. In Section \ref{main_sec}, a solver-free, multilevel spectral graph sparsification framework is described in detail.  Section
\ref{result_sec}  demonstrates extensive experiment results for a variety of real-world, large-scale   matrix and graph problems, which is followed by the conclusion of this work in Section \ref{conclusion}.

\section{Background}\label{background_sec}
% In this section, we introduce the basic concepts related to spectral graph sparsification and coarsening, which is followed by  discussions of the related works. 

\subsection{Graph Laplacians and Quadratic Forms}
%Let G = (V, E ), with N = |V |, be a simple undirected graph with adjacency A ∈ RN ×N, and data matrix X ∈ RN×p. which has p-dimensional real-valued vector representations for each node v ∈ V

Consider a weighted, undirected graph   $\mathcal{G}=(\mathcal{V},\mathcal{E}, \omega)$ with $\mathcal{|V|=N}$ and $\mathcal{|E|=M}$, where
$\mathcal{V}$ denotes a set of vertices, $\mathcal{N}$ denotes the number of vertices, $\mathcal{E}$ denotes a set of edges, $\mathcal{M}$ denotes the number of edges, and $\omega $ denotes a weight
function that assigns a positive weight to each edge. The adjacency matrix of graph $\mathcal{G}$ can be defined as  follows:
\begin{equation}\label{formula_adjacency}
\mathbf{\mathcal{A}_{\mathcal{G}}}(p,q)=\begin{cases}
\omega(p,q) & \text{ if } (p,q)\in \mathcal{E}\\

0 & \text{ if otherwise } .
\end{cases}
\end{equation}
%\begin{equation}\label{formula_adjacency}
%\mathbf{\mathcal{L}_{\mathcal{G}}}(p,q)=\begin{cases}
%-\omega(p,q) & \text{ if } (p,q)\in E \\
%\sum\limits_{(p,t)\in E}\omega(p,t) & \text{ if } (p=q) \\
%0 & \text{ if otherwise } .
%\end{cases}
%\end{equation} 
The Laplacian matrix can be computed by $\mathcal{L_G}=\mathcal{D_G}-\mathcal{A_G}$, where $\mathcal{D_G}$ is an diagonal matrix with elements $\mathcal{D}_{\mathcal{G}}(p,p)=\sum\limits_{t\ne p}\omega(p,t)$. For any  real vector $x\in {\mathbb{R} ^\mathcal{N}}$, the Laplacian quadratic form of graph $\mathcal{G}$ is defined as:
 $\mathbf{{x^\top}\mathcal{L}_{\mathcal{G}} x} = \sum\limits_{\left( {p,q} \right) \in E}
{{\omega({p,q})}{{\left( {x\left( p \right) - x\left( q \right)}
\right)}^2}}$. 
\subsection{Spectral Graph Sparsification}

% \textbf{Cut sparsifier}   was originally proposed by Benczur and Karger \cite{benczur1996approximating,benczur2015randomized}, which is defined as a re-weighted graph of the original graph such that the weights of every cut in the sparsifier could approximate the ones in the original graph correspondingly. 

\textbf{Spectral  sparsifier} was first introduced by Spielman and Teng \cite{spielman2011spectral}, which is a strictly stronger notation than the cut sparsifier \cite{benczur1996approximating,benczur2015randomized}. The spectral sparsifier is a weighted subgraph such that the difference of quadratic forms calculated by original graph and the sparsifier is bounded by $(1\pm\epsilon)$, where $\epsilon$ is a constant factor. Given an undirected graph with $\mathcal{N}$ vertices and $\mathcal{M}$ edges,     a nearly-linear time algorithm was introduced for building  $(1\pm \epsilon)$ spectral sparsifiers with $O(\mathcal{N} \log \mathcal{N}/\epsilon^2)$ edges in \cite{spielman2011graph}. Later, Batson, Spielman, and Srivastava  \cite{batson2012twice}  proposed the algorithm for constructing the sparsifier within $O(\mathcal{N}/\epsilon^2)$ edges. Recently, the state-of-the-art work is given by Lee and Sun \cite{Lee:2017} that computes a $(1\pm \epsilon)$ sparsifier with $O(q \mathcal{N}/\epsilon)$ edges  in nearly  linear time $O \left( \frac{q\mathcal{M}\mathcal{N}^{5/q}}{\epsilon^{4+4/q}} \right)$, where $q$ is an integer greater than $10$.
%   \begin{figure}[!htb]
%   \includegraphics[width=0.98\linewidth]{SpectralSparsification.pdf}
%   \caption{Spectral graph sparsification }\label{fig:ss}
% \end{figure}

Another metric for quantifying spectral similarity of two graphs has been proposed by
Spielman and Teng \cite{spielman2011graph}: the subgraph $\mathcal{P}=(\mathcal{V},E)$  is a $\sigma-$spectral sparsifier of the original graph  $\mathcal{G}$ if the following inequality holds for any $x\in {\mathbb{R} ^\mathcal{N}}$
\begin{equation}\label{eqn:ineq}
    \frac{1}{\sigma}{x^\top}\mathcal{L}_{\mathcal{G}} x \leq {x^\top}\mathcal{L}_{\mathcal{P}} x\leq \sigma {x^\top}\mathcal{L}_{\mathcal{G}} x,
\end{equation}
where the relative condition number is defined as $\kappa(\mathcal{L}_{\mathcal{G}},\mathcal{L}_{\mathcal{P}})\leq\sigma^2$. It indicates that a smaller relative condition number corresponds to a higher spectral similarity. 
% \textbf{Parallel solvers for SDD systems. }
% Peng and Spielman \cite{peng2014efficient}  presented a novel algebraic framework for solving symmetric, diagonally dominant  (SDD)  systems that runs in polylogarithmic time and nearly-linear work. It could use parallel sparsification algorithms for constructing parallel solvers. It motivates Koutis \cite{koutis2016simple} to generalize a parallel algorithm for spectral graph sparsification based on spanner and sampling.
\subsection{Spectral Graph Coarsening }
\textbf{Graph coarsening (reduction)} not
only reduces the number of edges but also aggregates nodes   to form smaller number of nodes for graph approximation.  It was at heuristics level until  Loukas, and Vandergheynst \cite{loukas2018spectrally,loukas2019graph}  developed a  theoretical   framework which guarantees that the spectral properties of coarsened graphs can approximate the original ones under some restricted circumstances.   

\section{SF-GRASS: Solver-Free  Graph Spectral Sparsification }\label{main_sec}
 The proposed \textbf{S}olver-\textbf{F}ree \textbf{Gra}ph \textbf{S}pectral \textbf{S}parsification (\textbf{SF-GRASS}) framework is built  upon a multilevel spectral graph coarsening scheme, which allows constructing   multilevel spectral sparsifiers in nearly-linear time. Given an undirected graph $\mathcal{G}=\mathcal{G}_0$, a series of reduced graphs $\mathcal{G}_1, \mathcal{G}_2,...,\mathcal{G}_{l_f}$ will be generated through a spectral coarsening procedure with the corresponding node sizes denoted by $\mathcal{N}_0, \mathcal{N}_1,...,\mathcal{N}_{l_f}$, where $\mathcal{N}_0>\mathcal{N}_1>...>\mathcal{N}_{l_f}$. Once the the coarsened graphs are constructed, the spectral sparsifier $\mathcal{P}_l$ of the coarsened graph $\mathcal{G}_l$ at  level $l$ will be extracted by the spectral perturbation approach introduced in \cite{feng2016spectral}, where $l=l_f,l_{f-1},...,0$, and $\mathcal{P}_0 = \mathcal{P}$ is defined as the spectral sparsifier for the original graph $\mathcal{G}$.  For the sake of simplicity, all the symbols used in this paper are summarized in Table \ref{tab:symbols}.
 
 \begin{center}
\begin {table*}
\caption {Summary of symbols used in the paper ($l=0,1,...,{l_f} , i=1,...,\mathcal{N}_l$).} \label{tab:symbols} 
\begin{tabular}{ c c|c c} 
 \hline
 symbols & description & symbols & description \\  \hline
 $\mathcal{G}_l=(\mathcal{V}_l,\mathcal{E}_l)$ & an undirected graph at level $l$ & $\mathcal{P}_l=(V_l,E_l)$ & the sparsifier of $\mathcal{G}_l$ \\ 
 $\mathcal{V}_l$ & node set at level $l$  & $\mathcal{V}_l$ & node set at level $l$    \\ 
 $\mathcal{E}_l$ & edge set of $\mathcal{G}_l$ & ${E}_l$ & edge set  of  $\mathcal{P}_l$\\
 ${\omega_{l}(p,q)}$ & edge weight of node $(p, q)$ for $\mathcal{G}_l$ & $\omega_l(p,q)$ & edge weight of node $(p, q)$ for $\mathcal{P}_l$\\ 
 $\mathcal{N}_l=|\mathcal{V}_l|$ & number of nodes & $\mathcal{N}_l$ & number of nodes \\
  $\mathcal{M}_l=|\mathcal{E}_l|$ & number of edges in $\mathcal{G}_l$ & ${M}_l=|E_l|$ & number of edges in $\mathcal{P}_l$ \\
 $\mathcal{L}_{\mathcal{G}_l}$ & Laplacian  of  graph $\mathcal{G}_l$ &$\mathcal{L}_{\mathcal{P}_l}$ & Laplacian of  graph $\mathcal{P}_l$\\[0.1cm] 
 $\mathcal{A}_{\mathcal{G}_l}$ & adjacency matrix of graph  $\mathcal{G}_l$& $\mathcal{A}_{\mathcal{P}_l}$ & adjacency matrix of  graph  $\mathcal{P}_l$\\[0.1cm]
 $\lambda_{l}^{(i)} $ & eigenvalues of  $\mathcal{L}_{\mathcal{G}_l}$&
 $\tilde{\lambda}_{l}^{(i)}$& eigenvalues of  $\mathcal{L}_{\mathcal{P}_l}$\\[0.1cm]
  $u_{l}^{(i)} $ & eigenvectors of  $\mathcal{L}_{\mathcal{G}_l}$&
 $\tilde{u}_{l}^{(i)} $& eigenvectors of  $\mathcal{L}_{\mathcal{P}_l} $\\ [0.1cm]
$S^{(i)}_{l-1}$  & \multicolumn{3}{c}{ node aggregation set at level $l-1$ with respect to the single node $i$ at level $l$} \\ 
 \hline
\end{tabular}
\end{table*}
\end{center}
% It can be shown that  every graph Laplacian  matrix is an symmetric diagonally dominant matrix (SDD) matrix, which also can  be considered as an admittance matrix of a resistor circuit network. For any  real vector $x\in {\mathbb{R} ^V}$, the Laplacian quadratic form of graph $G$ is defined as:
% $\mathbf{{x^\top}L_G x} = \sum\limits_{\left( {p,q} \right) \in E}
%{{\omega_{p,q}}{{\left( {x\left( p \right) - x\left( q \right)}
%\right)}^2}}$.
%Recently,  spectral methods for graph sparsification  and coarsening  \cite{spielman2011spectral,zhuo:dac16,zhuo:dac18,loukas2018spectrally,zhao2018nearly, zhao:dac19,loukas2019graph} have been developed based on spectral graph theory to allow dramatic reduction of the size of an undirected graph, while preserving the key spectral (structural) properties of the original graph, such as the first few Laplacian eigenvalues/eigenvectors, the cuts in the graph, effective resistances, and Laplacian quadratic forms.  

\subsection{Overview of Our Approach}

%In recent years, The size of data has been increasing rapidly in undirected graphs. We come up with very efficient algorithms for getting the sparsifier to approximate the original graph with the set of nodes and much fewer edges \cite{lee:edu}, \cite{feng2016spectral}, \cite{zhuo:dac18}, where the sparsifier aims to preserve spectral properties of the original one.   
Recent  research in graph signal processing (GSP) \cite{shuman2013emerging} shows that for undirected graphs the smaller eigenvalues and corresponding eigenvectors of its   Laplacian are associated to the global structure (long-range distances) of the underlying graph, while the higher eigenvalues and corresponding eigenvectors encode the local structure of the graph. Since spectral sparsification aims to approximate the first few eigenvalues and eigenvectors of the original Laplacian with the minimum number of edges, it can be regarded as a low-pass filter on graphs for removing redundant edges. Spectral sparsification usually involves two steps: the first step is to generate a low-stretch spanning tree (LSST) from the original graph using star or petal decompositions \cite{elkin2008lower,abraham2012}; the next step is to identify and recover spectrally-critical off-tree edges into the LSST  to drastically reduce the condition number, and thereby minimizing the spectral mismatch \cite{feng2020grass}. However, prior spectral sparsification methods \cite{spielman2011graph,feng2016spectral} usually require solving linear systems of equations with  Laplacian solvers, which can still be computationally challenging for large problems.

  \begin{figure}[!htb]
\minipage{0.5\textwidth}
  \includegraphics[width=0.94\linewidth]{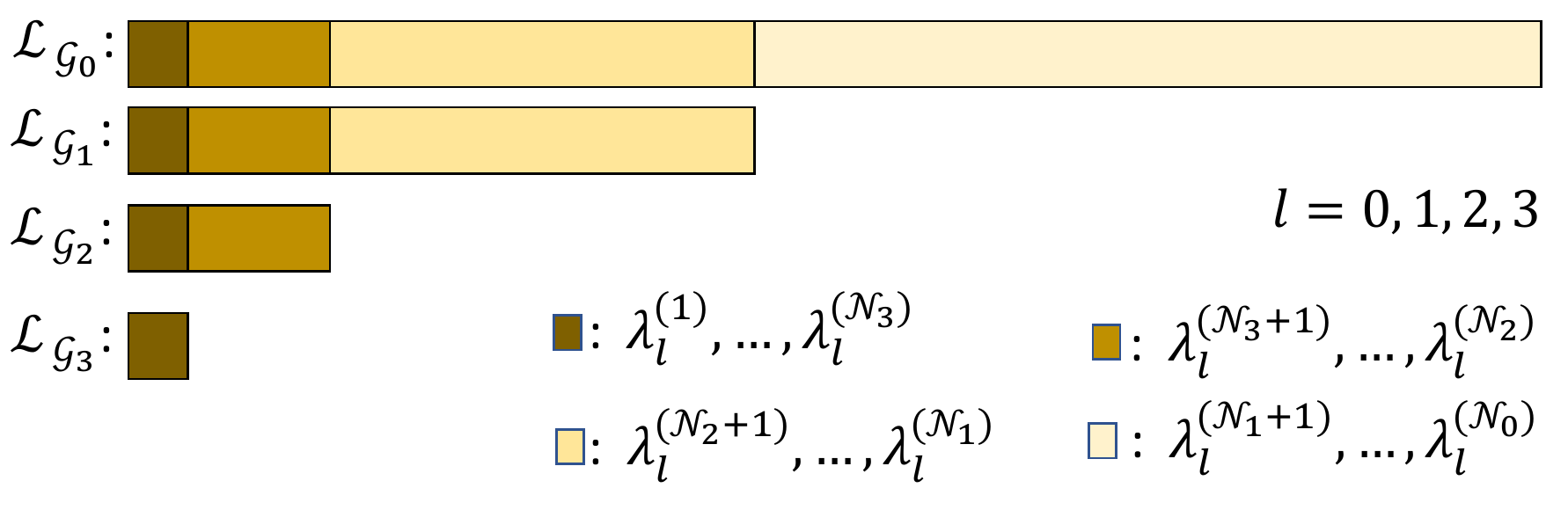}
  \caption{Eigenvalue distributions of  $\mathcal{L}_{\mathcal{G}_l}$ }\label{fig:orig_eig_dis}
\endminipage\hfill
\minipage{0.47\textwidth}
  \includegraphics[width=0.995\linewidth]{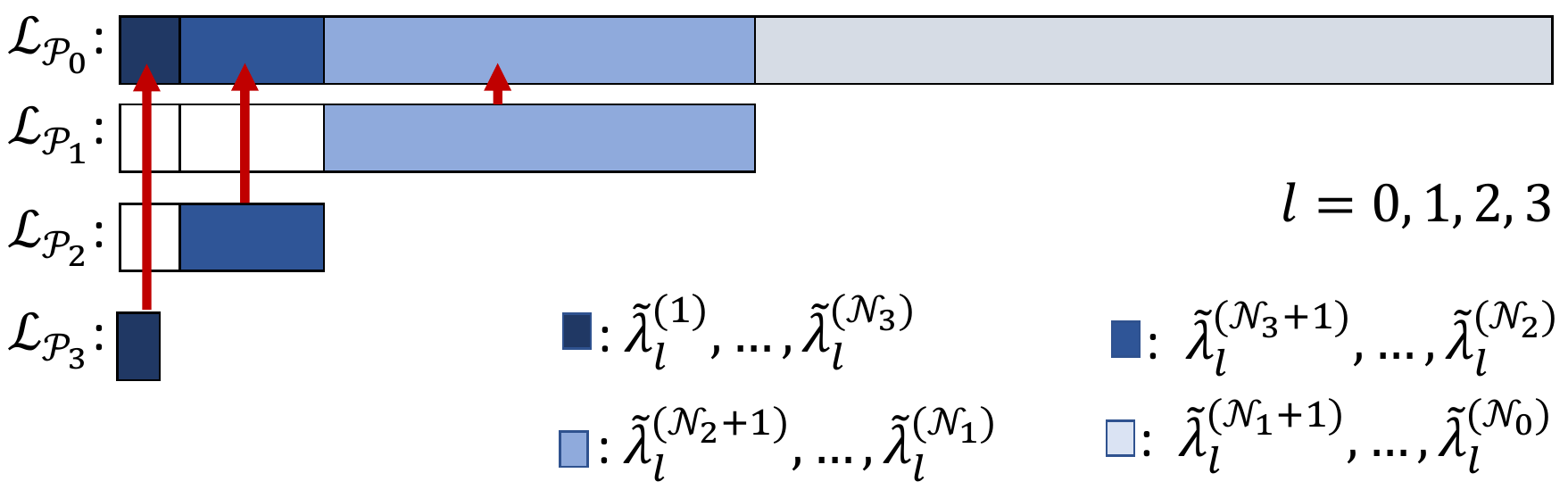}
  \caption{Eigenvalue distributions of  $\mathcal{L}_{\mathcal{P}_l}$ }\label{fig:spar_eig_dis}
\endminipage
\end{figure}

In this work, we propose a solver-free, multilevel spectral sparsification scheme to generate a hierarchy of increasingly smaller spectral sparsifiers. As aforementioned, given an undirected graph $\mathcal{G}_{0}= \mathcal{G}$, a series of coarsened  graphs $\mathcal{G}_1, \mathcal{G}_2,...,\mathcal{G}_{l_f}$ will be generated via the multilevel spectral graph coarsening scheme introduced in \cite{zhao:dac19}, where $\mathcal{G}_{l_f}$ denotes the coarsest graph. It can be shown that the Laplacian of   $\mathcal{G}_{l}$   can well preserve   the  low eigenvalues and eigenvectors   of the finer graphs $\mathcal{G}_{l-1}$, ... , $\mathcal{G}_{1}$, $\mathcal{G}_{0}$ \cite{loukas2018spectrally,loukas2019graph}. For example, Figure \ref{fig:orig_eig_dis} shows the eigenvalue distributions of the Laplacian matrices corresponding to four consecutive coarse-level   graphs, implying that the eigenvalues $\left(\lambda_l^{1}, ..., \lambda_l^{\mathcal{N}_3}\right)$ of  $\mathcal{L}_{\mathcal{G}_3}$  will approximately match  the smallest eigenvalues of $\mathcal{L}_{\mathcal{G}_{2}}$, $\mathcal{L}_{\mathcal{G}_{1}}$ and $\mathcal{L}_{\mathcal{G}_{0}}$. In other words, $\mathcal{L}_{\mathcal{G}_{3}}$ will always approximately preserve  the key spectral (structural) properties of $\mathcal{L}_{\mathcal{G}_{0}}$ after  coarsening. Similarly, eigenvalues $\left(\lambda_l^{\mathcal{N}_3+1}, ..., \lambda_l^{\mathcal{N}_2}\right)$ of $\mathcal{L}_{\mathcal{G}_2}$ will approximately match the first few eigenvalues of $\mathcal{L}_{\mathcal{G}_1}$ and $\mathcal{L}_{\mathcal{G}_0}$. Compared to  $\mathcal{G}_{3}$ and  $\mathcal{G}_{2}$,    $\mathcal{G}_{1}$ will retain more local information  of  $\mathcal{G}_{0}$ by approximately preserving the moderate to large eigenvalues of $\mathcal{L}_{\mathcal{G}_{0}}$. Consequently,   spectral coarsening is creating a hierarchy of smaller graphs that can be considered as \textbf{a cascade of low-pass graph filters} with gradually decreasing bandwidths: the finest graph always retains the highest bandwidth, whereas the coarsest graph only retains the lowest bandwidth. The theoretical proofs for the multilevel spectral preservation  via graph coarsening are provided in Section \ref{main:spectral_pre}.

Once the series of reduced graphs have been obtained via spectral coarsening, we will be able to effectively exploit them for extracting a hierarchy of ultra-sparse spectral sparsifiers. In the following, we show detailed steps for constructing spectral sparsifiers $\mathcal{P}_{l}$ at each level $l=l_f, ..., 1, 0$, such that each $\mathcal{P}_{l}$ will be spectrally-similar to  $\mathcal{G}_{l}$. Unlike the spectral coarsening step that starts at the finest-level (original) graph, SF-GRASS will start from the coarsest-level graph $\mathcal{G}_{l_f}$ and aims to approximate eigenvalues  and eigenvectors (in an ascending order) through a stratified scheme: when   $\mathcal{G}_{l_f}$ is sufficiently small, we can always efficiently extract the spectral sparsifier $\mathcal{P}_{l_f}$ for level ${l_f}$, leading to good approximation of the first few eigenvalues (eigenvectors); then we will map $\mathcal{P}_{l_f}$ to the finer level to facilitate the construction of the next-level sparsifier $\mathcal{P}_{l_f-1}$ so that higher eigenvalues (eigenvectors) can be approximated. The proposed approach SF-GRASS strives to incrementally construct a series of  increasingly finer spectral sparsifiers, as shown in Figure \ref{fig:spar_eig_dis}. After iteratively applying the above procedure for all levels, the spectral sparsifier $\mathcal{P}_0$ for the original graph can be efficiently constructed to well preserve the key spectral properties of  $\mathcal{G}_{0}$.
%In Figure \ref{fig:spar_eig_dis}, it shows the eigenvalue distribution of sparsifiers $\mathcal{L}_{\mathcal{P}_{l}}$. $\mathcal{L}_{\mathcal{P}_3}$ could preserve the smallest set of eigenvalues of $\mathcal{L}_{\mathcal{P}_{0}}$. 

% The rest of this section is organized as follows. Section  \ref{subsetion:Spectral Coarsening} introduces a  spectral graph coarsening method based on local spectral embedding. Section \ref{main:spectral_pre} proves the preservation of key spectral properties on coarse-level graphs obtained by spectral coarsening. Section \ref{subsetion:backward mapping} shows the detailed spectral sparsifier mapping process. Section \ref{subsection:seeking} introduces a highly-efficient scheme for identifying spectrally-critical  off-subgraph edges for further improving spectral approximations in spectral sparsifiers. Section \ref{main:complexity} provides the complete SF-GRASS algorithm flow and complexity analysis.
  \begin{figure*}[!htb]
\minipage{0.33\textwidth}
  \includegraphics[width=\linewidth]{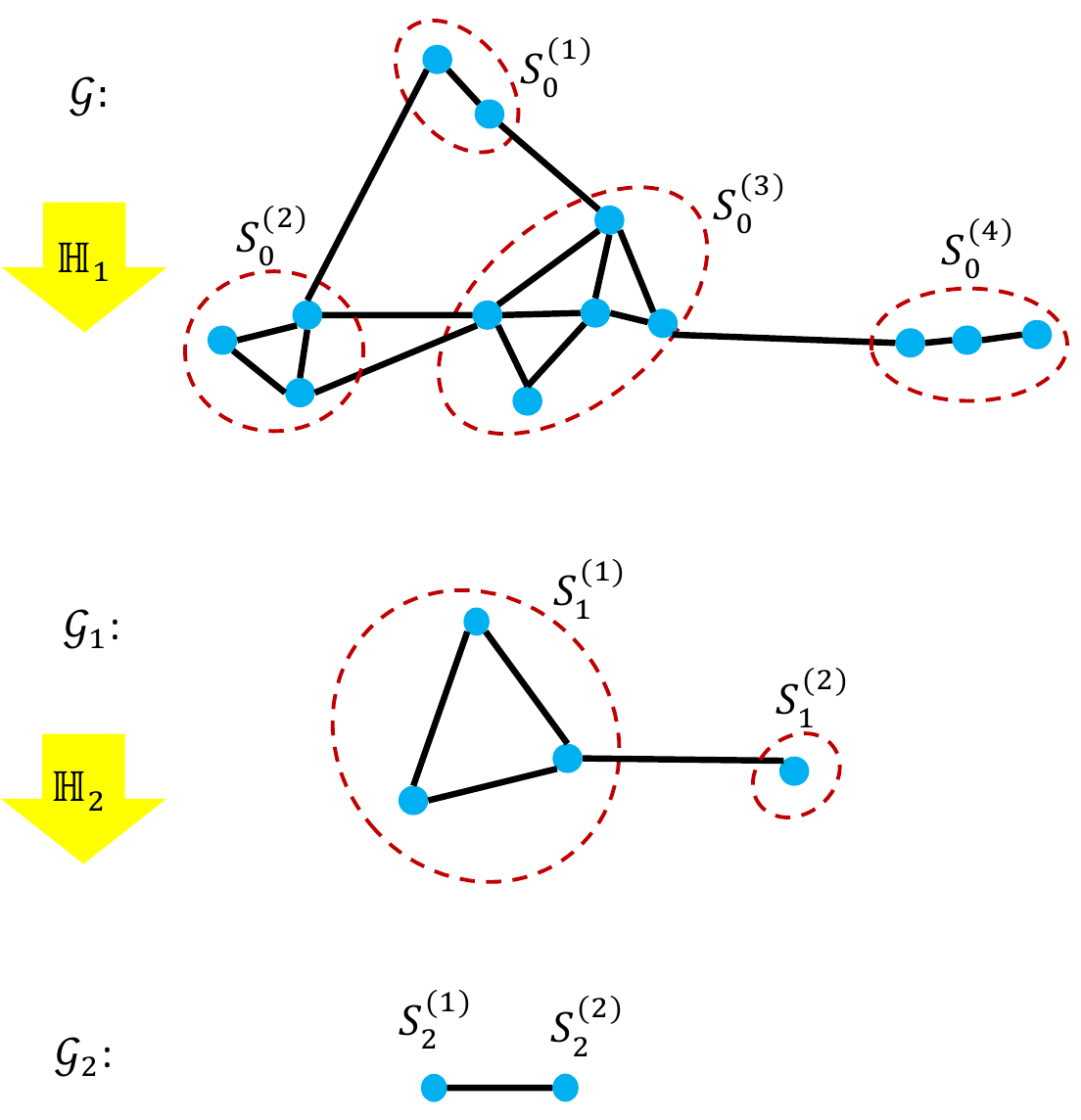}
  \caption{Graph spectral coarsening via local embedding}\label{fig:coarsening}
\endminipage\hfill
\minipage{0.32\textwidth}
  \includegraphics[width=\linewidth]{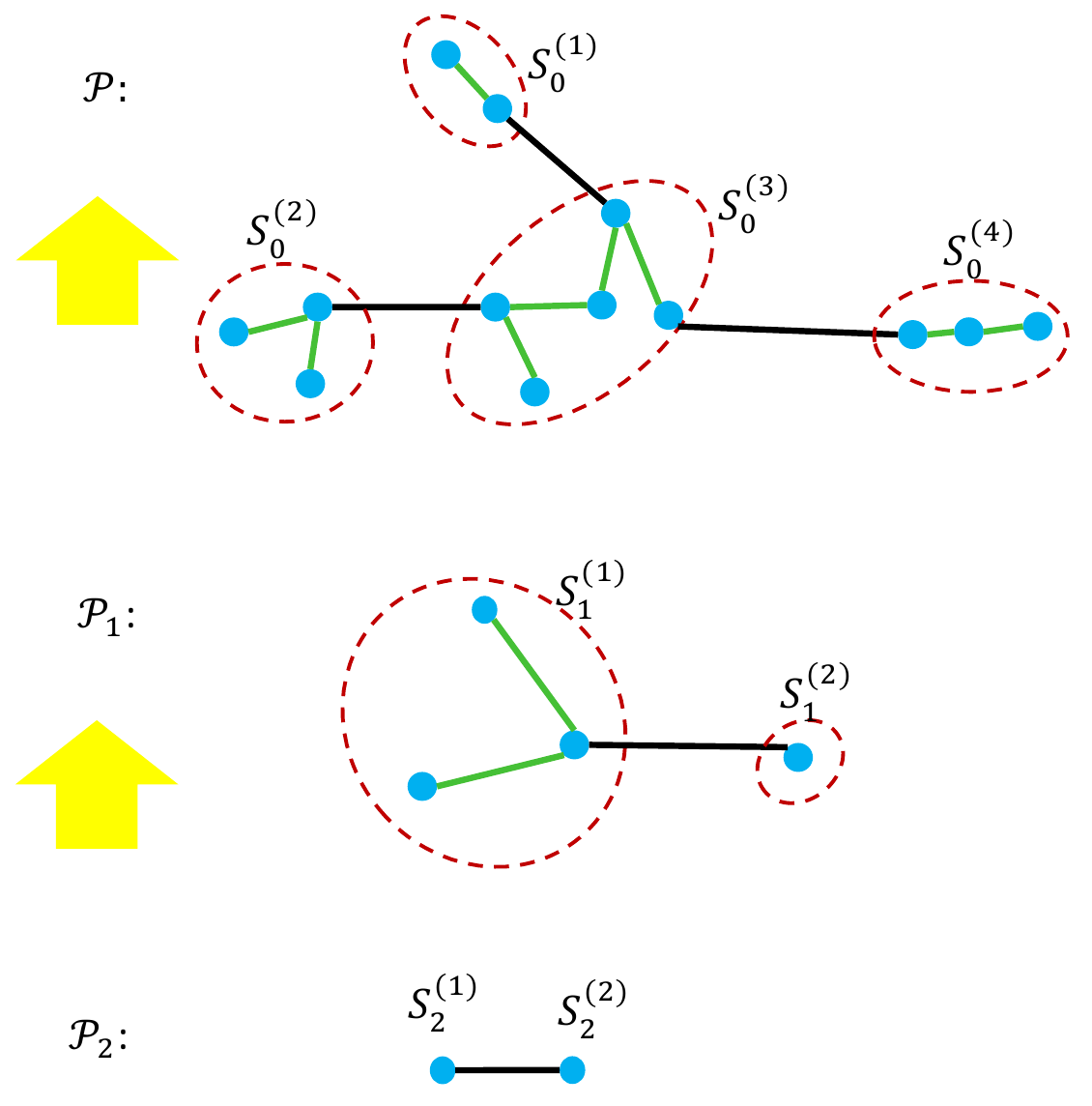}
  \caption{Sparsifier backward mapping }\label{fig:backward}
\endminipage\hfill
\minipage{0.32\textwidth}%
  \includegraphics[width=\linewidth]{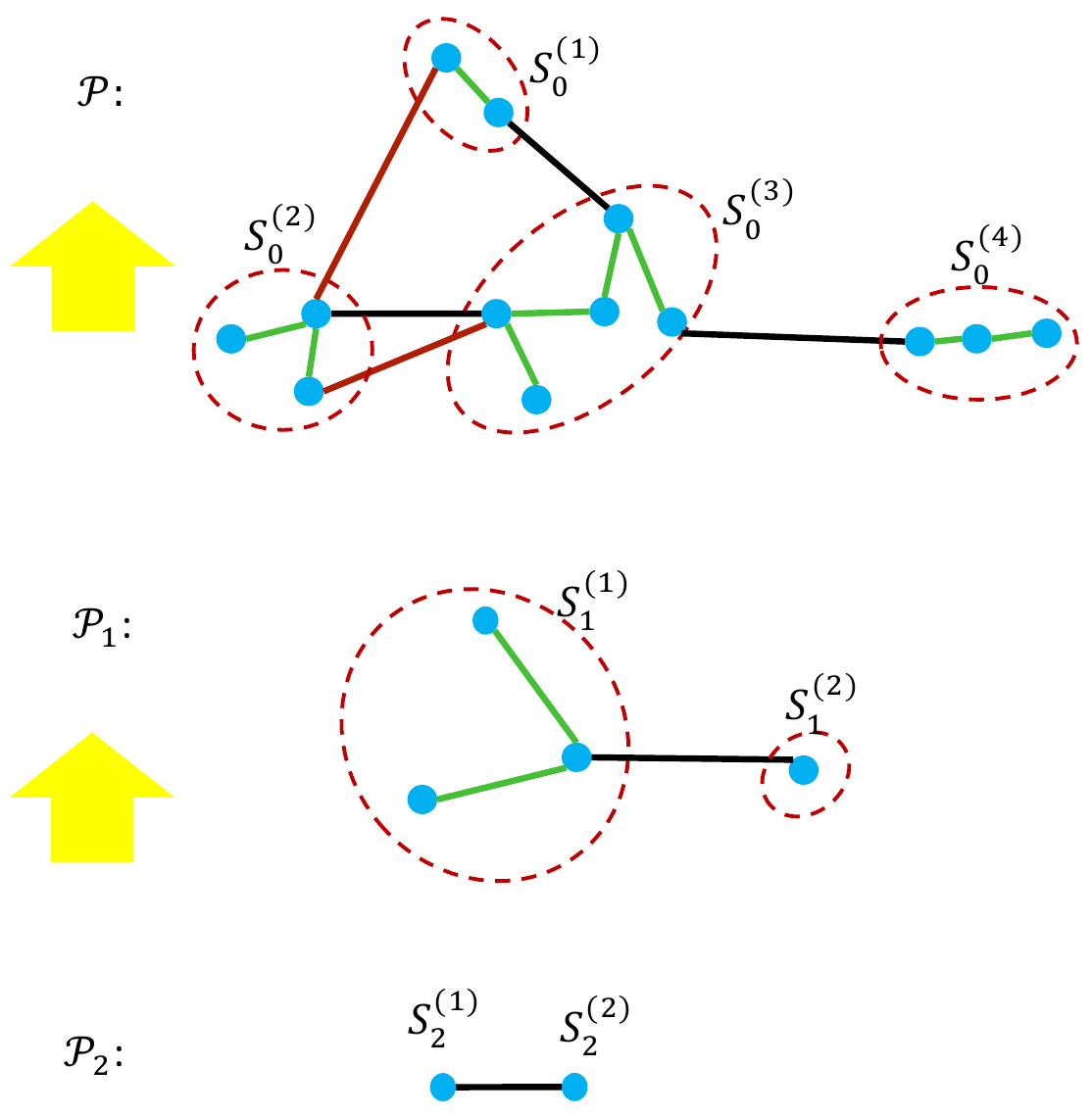}
  \caption{ Spectrally-critical edge identification }\label{fig:emebedding}
\endminipage
\end{figure*}
\subsection{ Spectral Coarsening  via Local Embedding }\label{subsetion:Spectral Coarsening}

As shown in Figure \ref{fig:coarsening}, an induced subgraph $\mathcal{F}^{(i)}_{l-1}$ can be   constructed with the node aggregation  set $\mathcal{S}^{(i)}_{l-1}$ and the edge set  $\mathcal{E}_{l-1} (\mathcal{S}^{(i)}_{l-1})$ that includes  edges $(p,q)$ in $\mathcal{E}_{l-1}$ with both of the nodes p and q included in the set $\mathcal{S}^{(i)}_{l-1}$. The induced subgraphs are strongly-connected components in $\mathcal{G}_{l-1}$, which will be aggregated into a single node of the  coarser graph.  Next, we   create a node-mapping matrix $\mathbb{H}_{l}$ that allows constructing  $\mathcal{G}_{l}$ given   the finer graph  $\mathcal{G}_{l-1}$ with the following equation:
\begin{equation}
  \mathcal{L}_{\mathcal{G}_{l}}:=\mathbb{H}_{l}^{\mp}\mathcal{L}_{\mathcal{G}_{l-1}}\mathbb{H}_{l}^{+}, \textbf{ and }       x_{l}:=\mathbb{H}_{l}x_{l-1}, ~~  \textbf{for}~ l=1,2,...,l_f,
\end{equation}
where $\mathbb{H}_{l} \in \mathbb{R}^{\mathcal{N}_{l}\times \mathcal{N}_{l-1}}$,   $x_{l}\in \mathbb{R}^{\mathcal{N}_l\times 1}$, and $\mathbb{H}_{l}^\top,\mathbb{H}_{l}^+,\mathbb{H}_{l}^\mp$ denote the transpose, pseudoinverse, and transposed pseudoinverse  of $\mathbb{H}_{l}$, respectively.  $\mathbb{H}_{l},\mathbb{H}_{l}^+$ can be created as follows \cite{loukas2019graph}:

% \noindent\begin{minipage}{.5\linewidth}
% \begin{equation}
%     \mathbb{H}_{l}(i,p)=\begin{cases}
% \frac{1}{|\mathcal{S}^{(i)}_{l-1}|} & \text{ if  node p}  \in  \mathcal{S}_{l-1}^{(i)} \\
% 0 & \text{ if otherwise } .
% \end{cases} 
% \end{equation}
% \end{minipage}%
% \begin{minipage}{.5\linewidth}
% \begin{equation}
% [\mathbb{H}^+]_{l}(p,i)=\begin{cases}
% 1 & \text{ if  node p}  \in  \mathcal{S}_{l-1}^{(i)} \\
% 0 & \text{ if otherwise } 
% \end{cases}
% \end{equation}
% \end{minipage}

\begin{equation}
    \mathbb{H}_{l}(i,p)=\begin{cases}
\frac{1}{|\mathcal{S}^{(i)}_{l-1}|} & \text{ if  node p}  \in  \mathcal{S}_{l-1}^{(i)} \\
0 & \text{ if otherwise } .
\end{cases} 
\end{equation}

\begin{equation}
\mathbb{H}^+_{l}(p,i)=\begin{cases}
1 & \text{ if  node p}  \in  \mathcal{S}_{l-1}^{(i)} \\
0 & \text{ if otherwise. } 
\end{cases}
\end{equation}

%in Procedure A. 
%\begin{center}
   
%\fbox{\begin{minipage}{35em}  
%Given the undirected graph $\mathcal{G}_l$ and a vector $x_l$ at the $l-th$ level, we can produce the next coarser graph $\mathcal{G}_{l+1}$ by the following equation.
%\begin{equation}
%  \mathcal{L}_{\mathcal{G}_{l+1}}:=\mathbb{H}_l^{\mp}\mathcal{L}_{\mathcal{G}_l}\mathbb{H}_l^{+} \textit{ and }       x_{l+1}:=\mathbb{H}_lx_l \qquad  l=0,1,..l_f-1
%\end{equation}

%%Where $\mathbb{H}_i \in \mathcal{R}^{\mathrm{N}_{i+1}\times \mathcal{N}_i}$ have more columns than rows. $\mathbb{H}_i^{+}$ means the pseudoinverse of $\mathbb{H}_i$ and $\mathbb{H}_i^{\mp}$ means the transpose of the pseudoinverse of $\mathbb{H}_i$.

%\end{minipage}}
%\end{center}
%The details about spectral properties and cut preserving between $\mathcal{G}_{l+1}$ and $\mathcal{G}_{l}$ are proved and informed in \cite{loukas2019graph}. 
When creating a coarsening framework, the core task is to cluster the graph into   aggregation sets so that we can define matrix $\mathbb{H}_l$. To preserve important spectral properties (e.g., the first few eigenvalues and eigenvectors of the graph Laplacian) on the coarsened graphs, one naive approach is to embed the original graph into a $K$-dimensional space using the first $K$ nontrivial Laplacian eigenvectors. Then, the nodes that are close to each other in the embedding space can be aggregated for forming a coarser graph.  However, such a scheme requires calculating the Laplacian eigenvectors, which will be extremely expensive for large graphs. 

To achieve good efficiency, SF-GRASS leverages a linear-time local spectral embedding  scheme based on low-pass filtering of random graph signals \cite{zhao:dac19,deng2019graphzoom}.  
 Let $X_l=[x_{l}^{(1)}, x_{l}^{(2)},...,x_{l}^{(K)}]$, where  $x_{l}^{(i)} \in \mathbb{R}^{\mathcal{N}_l\times 1}$ denote the test vectors computed by applying a few steps of Gaussian-Seidel relaxations for solving
the linear system of equations $\mathcal{L}_{\mathcal{G}_{l}} x_{l}^{(i)} =0 $ for i = 1, ..., K
with $K$ initial random vectors that are orthogonal to the all-one
vector \cite{livne2012lean}. The above smoothing procedure can be regarded as a low-pass filtering process applied to $K$ random graph signals. The resultant $K$ smoothed test vectors  will consist of linear combinations of the first few Laplacian eigenvectors, and thus can be subsequently leveraged for spectral graph embedding.

% The spectral node proximity metric based on algebraic distance could be calculated at a high level of efficiency  in the low-dimensional embedding space.  The spectral node affinity $a_{i,j}$  according to node p and q is defined as follows (\cite{livne2012lean,chen2011parallel})
% \begin{equation}\label{eqn: algebraic distance}
% a_{p,q}= \frac{(X_l(p,:),X_l(q,:))^2}{\|X_l(p,:)\|^2_{L_2}\|X_l(q,:)\|^2_{L_2}}  
% \end{equation}

% Where $(X_l(p,:),X_l(q,:))$ is the inner product of $X_l(p,:)$ and $X_l(q,:)$.  A larger $a_{p,a}$ value indicates a closer algebraic distance (stronger spectral similarity) between nodes p and q. Therefore, the graph nodes in the same aggregation set will be recognized and could be merged  to form the coarse-level nodes and subsequently the reduced graph. 

Since  modern {graph signal processing (GSP) based filtering functions  are dominated by SpMV operations} that  are massively-parallel-friendly, the  local spectral embedding scheme can be effectively accelerated on modern parallel computing platforms, such as CPUs, GPUs, and FPGAs \cite{steinberger2017globally,hong2018efficient}. 

\subsection{Spectral Similarity Between Coarse Graphs}\label{main:spectral_pre}
After finding the $\mathbb{H}_{l}$, will Eq (\ref{eqn:ineq}) still hold between  $\mathcal{G}_{l}$ and $\mathcal{G}_{l-1}$?  If yes, how does the smaller graph $\mathcal{G}_{l}$ spectrally preserve the finer graph $\mathcal{G}_{l-1}$?  How will  $\mathbb{H}_l$ affect the spectral properties of $\mathcal{G}_{l}$? For the above questions, we provide detailed explanation through  the following comprehensive theoretical analysis. Let $\lambda_{l}^{(1)},\lambda_{l}^{(2)},..,\lambda_{l}^{(\mathcal{N}_l)}$, and $u_{l}^{(1)}, u_{l}^{(2)},...,u_{l}^{(\mathcal{N}_l)}$ denote the non-decreasing eigenvalues  and their corresponding eigenvectors for    $\mathcal{L}_{\mathcal{G}_l}$.
The \textbf{restricted spectral similarity}  \cite{loukas2019graph} is defined  as follows
\begin{equation}
   \frac{1}{\sigma_{l-1}}\|x_{l-1}\|_{\mathcal{L}_{\mathcal{G}_{l-1}}}\leq \|x_{l}\|_{\mathcal{L}_{\mathcal{G}_{l}}}\leq \sigma_{l-1}\|x_{l-1}\|_{\mathcal{L}_{\mathcal{G}_{l-1}}},~ \forall  x_{l-1} \in U^k_{l-1},
\end{equation}
where  $U^k_{l-1}=\left[u_{l-1}^{(1)},u_{l-1}^{(2)},..., u_{l-1}^{(k)}\right]$ includes the first $k$ eigenvectors   of $\mathcal{L}_{\mathcal{G}_{l-1}}$.   The restricted spectral similarity can also be denoted as the $(U^k_{l-1}, \sigma_{l-1})$-spectral similarity.
If   $\mathcal{L}_{\mathcal{G}_{l-1}}$ and  $\mathcal{L}_{\mathcal{G}_l}$ are $(U^k_{l-1}, \sigma_{l-1})$-similar,  we have
  \begin{equation}
     \gamma_1 \lambda_{l-1}^{(i)}\leq
  \lambda_{l}^{(i)}\leq\gamma_2 {\small{\frac{(1+\epsilon)^2}{1-\tau\epsilon^2}}} \lambda_{l-1}^{(i)} \;, \;\;i = 1\;, \cdots\;,\; \mathcal{N}_l
 \end{equation}
  where $  \tau=\lambda^{(k)}_{l-1}/\lambda^{(2)}_{l-1}, \epsilon=(\sigma_{l-1}^2-1)/(\sigma_{l-1}^2+1)$ and $\sigma_{l-1}\leq (\frac{1+\sqrt{\tau}}{1-\sqrt{\tau}})^{\frac{1}{2}}$.  $\gamma_1,\gamma_2$ will be the smallest and largest eigenvalues of $(\mathbb{H}_l\mathbb{H}_l^\top)^{-1}$. Therefore, the spectral similarity between $\lambda_{l-1}^{(i)}$ and $\lambda_{l}^{(i)}$ can be controlled by $\sigma_{l-1}$. The canonical angles between the principal eigenspace  of $\mathcal{L}_{\mathcal{G}_{l-1}}$ and $\mathcal{L}_{\mathcal{G}_l}$  are defined as  follows:
  \begin{equation}
      \Theta(U^k_{l-1},\mathbb{H}_{l}^\top U^k_{l} )=\textbf{arccos}(U^{k\top}_{l-1}\mathbb{H}_{l}^\top U^{k}_{l} U^{k\top}_{l} \mathbb{H}_{l}U^k_{l-1})^{-\frac{1}{2}}.
  \end{equation}
  Consequently, a smaller canonical angles implies a higher similarity between two eigenspaces. 
 
%\end{enumerate}

%\begin{enumerate}[label= \textbf{Property} \arabic*.,itemindent=*]
 % \item For matrix  %$\mathcal{L}_{\mathcal{G}_l}$ with full-row rank,  we have
 % \begin{equation}
 %%     \gamma_1 \lambda_{l}^{(i)}\leq\lambda_{l+1}^{(i)}\leq\gamma_2\lambda_{l}^{(i+\mathcal{N}_l-\mathcal{N}_{l+1})}
 % \end{equation}
  %
 % where $i=1,2,...\mathcal{N}_{l+%1}$ and $\gamma_1,\gamma_2$ will be smallest and largest eigenvalue of $(\mathbb{H}_l\mathbb{H}_l^\top)^{-1}$ 

 % \item    For any vector $x_{l+1}$, we have
 % \begin{equation}
 %     x_{l+1}^\top \mathcal{L}_{\mathcal{G}_{l+1}}x_{l+1}=x_{l}^\top \Pi_l\mathcal{L}_{\mathcal{G}_{l}}\Pi_i x_{l}=x_{l}^\top  \mathcal{L}_{\mathcal{G}_{l}}x_{l} \textit{ and } x_{l+1}=\Pi_l x_{l}
  %\end{equation}
 
%\end{enumerate}

\subsection{Sparsifier Backward Mapping }\label{subsetion:backward mapping}

%Once the  smallest reduced graph $\mathcal{G}_{l_f}$ is obtained,  let $\mathcal{G}_{l_f}=\mathcal{P}_{l_f}$.
%We aim to find the undirected graph $\mathcal{G}_{l-1}$'s sparsifier $\mathcal{P}_{l-1}$ from  $\mathcal{P}_{l}$ and $\mathbb{H}_{l}$, where $l={l_f},{l_f}-1,...,1 $. The nodes in graph  $\mathcal{G}_{l}$ represent the clusters in the  $\mathcal{G}_{l-1}$ and  the edges in graph  $\mathcal{G}_{l}$ mean the outer-cluster edges in the $\mathcal{G}_{l-1}$. In addition, $\mathcal{P}_{l-1}$ is the sparsifier of  graph $\mathcal{G}_{l-1}$.

We aim to iteratively find spectral sparsifiers  for achieving   desired spectral similarity or  relative condition numbers   $\kappa(\mathcal{L}_{\mathcal{G}},\mathcal{L}_{\mathcal{P}})=\sigma_l^2$, $l=l_f$,...,0. To this end, we will  first extract an LSST, and subsequently, add extra off-tree edges to form the sparsifier  at the coarsest level. Next, sparsifiers at finer levels can be obtained by iteratively mapping the coarser sparsifiers via the procedures illustrated in Figures \ref{fig:coarsening} and \ref{fig:backward}, where $\mathcal{P}$, $\mathcal{P}_{1}$, $\mathcal{P}_{2}$ denote the sparsifiers of  $\mathcal{G}$,  $\mathcal{G}_{1}$, $\mathcal{G}_{2}$, respectively.  $\mathcal{P}_{l}$ and $\mathcal{G}_{l}$  share the same aggregation sets $S_{l-1}^{(i)}$.  $\mathcal{P}_{l}$ is the series of coarsened graphs for $\mathcal{P}$  where $l=1, 2$. 
In the following, we describe how to map two types of edges in the proposed sparsifier backward  mapping procedure:
\begin{itemize}
    \item \textbf{ Inner-cluster Edges.}
    We can conveniently locate all the inner-cluster nodes and edges within each aggregation set  according to   $\mathbb{H}_{l}$ and $\mathcal{G}_{l-1}$. Since each of  the aggregation sets is a strongly-connected component, we can extract an LSST for each  aggregation set (highlighted by the red dash line in Figure \ref{fig:backward}).  Since  each aggregation set size is pretty small, LSSTs can be well approximated using all-pairs shortest-path trees or maximum spanning trees (MSTs). 
    
    \item \textbf{ Inter-cluster Edges.}
We could get all the inter-cluster edges between these aggregation sets in the graph $\mathcal{G}_{l-1}$  and only keep the edges with the largest weights in $\mathcal{P}_{l-1}$.  As a result, all the aggregation sets will be connected through   inter-cluster edges in $\mathcal{P}_{l-1}$, forming a good  spectral sparsifier for $\mathcal{G}_{l-1}$ (as shown in Figure \ref{fig:backward}). 

\end{itemize}{}

\subsection{Spectrally-Critical  Edges Identification}
\label{subsection:seeking}
To further improve the spectral approximation in sparsifiers, additional spectrally-critical off-tree edges need to be identified and added into the latest sparsifiers. Specifically,
$O(\frac{\mathcal{M}_l \log \log \mathcal{N}_l}{\sigma^2})$   spectrally-critical off-tree edges  need to be  added into LSSTs to obtain a $\sigma$-similar spectral sparsifier $\mathcal{P}_{l}$ for $\mathcal{G}_{l}$. Let   $\tilde{u}_{l}^{(i)} $ denote the $i$-th eigenvector of $\mathcal{L}_{\mathcal{P}_l}$ corresponding to the $i$-th eigenvalue ${\tilde{\lambda}_{l}^{(i)}}$ that satisfies:
\begin{equation}\label{formula_eig_perturb0}
\mathcal{L}_{\mathcal{P}_l}\tilde{u}_{l}^{(i)} =\tilde{\lambda}_{l}^{(i)} \tilde{u}_{l}^{(i)} ,
\end{equation}
then we have the following eigenvalue perturbation analysis:
\begin{equation}\label{formula_eig_perturb1}
\left( {\mathcal{L}_{\mathcal{P}_l}+ \delta \mathcal{L}_{\mathcal{P}_l}} \right)\left( {{\tilde{u}_{l}^{(i)} } + \delta {\tilde{u}_{l}^{(i)} }} \right) = \left( {{\tilde{\lambda}_{l}^{(i)}} + \delta {\tilde{\lambda}_{l}^{(i)}}} \right)\left( {{\tilde{u}_{l}^{(i)} } + \delta {\tilde{u}_{l}^{(i)} }} \right),
\end{equation}
where a perturbation $\delta \mathcal{L}_{\mathcal{P}_l}$ that includes a new edge connection  is applied to $\mathcal{L}_{\mathcal{P}_l}$, resulting in perturbed eigenvalues and eigenvectors  ${\tilde{\lambda}_{l}^{(i)}} + \delta {\tilde{\lambda}_{l}^{(i)}}$ and ${\tilde{u}_{l}^{(i)} } + \delta {\tilde{u}_{l}^{(i)} }$ for $i=1,...,\mathcal{N}_l$, respectively. Keeping only the first-order terms leads to:
\begin{equation}\label{formula_eig_perturb1_first_order}
 {\mathcal{L}_{\mathcal{P}_l}}\delta {\tilde{u}_{l}^{(i)} } + {\delta \mathcal{L}_{\mathcal{P}_l}}{\tilde{u}_{l}^{(i)} } = {{\tilde{\lambda}_{l}^{(i)}}{\delta {\tilde{u}_{l}^{(i)} }} + \delta {\tilde{\lambda}_{l}^{(i)}}}  {{\tilde{u}_{l}^{(i)} } }.
\end{equation}
Expressing $\delta \tilde{u}_{l}^{(i)} $ in terms of the original eigenvectors $\tilde{u}_{l}^{(j)} $  for $j=1,...,\mathcal{N}_l$ leads to:

\begin{equation}\label{delta u_i}
{\delta {\tilde{u}_{l}^{(i)} }} = \sum\limits_{ j = 1}^{\mathcal{N}_l} {{\alpha _j}{\tilde{u}_{l}^{(j)} }}.
\end{equation}
Substituting (\ref{delta u_i}) into (\ref{formula_eig_perturb1_first_order}) leads to:
\begin{equation}\label{formula_eig_perturb1_first_order_expand}
 {\mathcal{L}_{\mathcal{P}_l}}\sum\limits_{j = 1}^{\mathcal{N}_l} {{\alpha _j}{\tilde{u}_{l}^{(j)} }} + {\delta \mathcal{L}_{\mathcal{P}_l}}{\tilde{u}_{l}^{(i)} } = {{\tilde{\lambda}_{l}^{(i)}}\sum\limits_{j = 1}^{\mathcal{N}_l} {{\alpha _j}{\tilde{u}_{l}^{(j)} }} + \delta {\tilde{\lambda}_{l}^{(i)}}}  {{\tilde{u}_{l}^{(i)} } }.
\end{equation}
Multiplying ${\tilde{u}_{l}^{(i)\top}}$ to both sides of (\ref{formula_eig_perturb1_first_order_expand}) results in:

\begin{equation}\label{formula_eig_perturb1_first_order_multiply}
\begin{split}
    &{{\tilde{u}_{l}^{(i)\top}} }{\mathcal{L}_{\mathcal{P}_l}}\sum\limits_{j = 1}^{\mathcal{N}_l} {\alpha _j}{\tilde{u}_{l}^{(j)} } + {\tilde{u}_{l}^{(i)\top}} \delta \mathcal{L}_{\mathcal{P}_l}\tilde{u}_{l}^{(i)}\\
    & = {{\tilde{\lambda}_{l}^{(i)}}{\tilde{u}_{l}^{(i)\top}} \sum\limits_{j = 1}^{\mathcal{N}_l} {{\alpha _j}{\tilde{u}_{l}^{(j)} }} + \delta {\tilde{\lambda}_{l}^{(i)}}}{\tilde{u}_{l}^{(i)\top}} {\tilde{u}_{l}^{(i)} }.
\end{split}
\end{equation}
Since $\tilde{u}_{l}^{(i)} $   for $i=1,...,\mathcal{N}_l$ are unit-length, mutually-orthogonal eigenvectors, we have:
\begin{equation}\label{formula_eig_perturb1_first_order_huajian}
\begin{split}
    & {\tilde{u}_{l}^{(i)\top}} {\mathcal{L}_{\mathcal{P}_l}}\sum\limits_{j = 1}^{\mathcal{N}_l} {{\alpha _j}{\tilde{u}_{l}^{(j)} }}  = {\alpha _i}{\tilde{u}_{l}^{(i)\top}} {\mathcal{L}_{\mathcal{P}_l}}{{\tilde{u}}}_{l}^{(i)} ,\\
    &{\tilde{\lambda}_{l}^{(i)}}{\tilde{u}_{l}^{(i)\top}} \sum\limits_{j = 1}^{\mathcal{N}_l}{\alpha _j}{\tilde{u}_{l}^{(j)} }={\alpha _i}{\tilde{u}_{l}^{(i)\top}} {\tilde{\lambda}_{l}^{(i)}}{\tilde{u}_{l}^{(i)} }.
\end{split}
\end{equation}
 Then the eigenvalue perturbation due to ${\delta \mathcal{L}_{\mathcal{P}_l}}$   is given by:

\begin{equation}\label{formula_eig_perturb1_conclusion}
 \delta {\tilde{\lambda}_{l}^{(i)}} = \omega({p,q})\left( {{{\tilde{u}_{l}^{(i)\top} }e_{p,q}} } \right)^2.
\end{equation}
Therefore, if an edge $(p,q)$ has a large $\omega(p,q)\left( {{{{\tilde{u}_{l}^{(i)\top}} }e_{p,q}}  } \right)^2$ value, it is considered  \textbf{spectrally critical} to $\tilde{\lambda}_{l}^{(i)}$. In other words, including this edge into the latest sparsifier will significantly perturb the Laplacian eigenvalue $\tilde{\lambda}_{l}^{(i)}$ and eigenvector $\tilde{u}_{l}^{(i)}$. Construct a  subspace matrix for $K$-dimensional spectral graph embedding using the first $K$    Laplacian eigenvectors as follows:
\begin{equation}\label{subspace}
U=\left[\tilde{u}_{l}^{(1)}, \tilde{u}_{l}^{(2)},..., \tilde{u}_{l}^{(K)}\right], 
\end{equation}
then  the overall $K$-eigenvalue  perturbation $\Delta_K$ becomes
\begin{equation}\label{formula_eig_perturb1_conclusion2}
\Delta_K=\sum\limits_{i = 1}^{{K}}  \delta {\tilde{\lambda}_{l}^{(i)}} = \omega({p,q})\left( U^\top e_{p,q} \right)^2,
\end{equation}
which is similar to the effective-resistance edge sampling probability \cite{spielman2011graph} computed by $P=\omega({p,q}) R^{eff}_{p,q}$ when $K=\mathcal{N}_l$, considering the close connection between the spectral embedding distance $\left( U^\top e_{p,q} \right)^2$ and the effective resistance distance  computed by 
\begin{equation}\label{formula_Reff}
R^{eff}_{p,q}=\left( U_{eff}^\top e_{p,q} \right)^2,~~~\text{where~~~} U_{eff}=\left[\frac{\tilde{u}_{l}^{(1)}}{\sqrt{\tilde{\lambda}_{l}^{(i)}}},..., \frac{\tilde{u}_{l}^{(\mathcal{N}_l)}}{\sqrt{\tilde{\lambda}_{l}^{(\mathcal{N}_l)}}}\right]. 
\end{equation}
 Instead of computing   exact Laplacian eigenvectors for identifying spectrally-critical edges, we will adopt the local spectral embedding approach described in Section \ref{subsetion:Spectral Coarsening} for approximately computing the spectral embedding distance $\left( U^\top e_{p,q} \right)^2$. Consequently, we can compute  spectral criticalities for all candidate edges in linear time using only   a few times of sparse-matrix-vector multiplications.

\subsection{Algorithm Flow and Complexity }\label{main:complexity}
\begin{algorithm}[!htbp]
\small { \caption{ ${\mathcal{P}}$ = SF-GRASS(${\mathcal{G}}$, $\sigma$)} \label{alg:directed_graph_spar}

%  \algsetup{indent=1em, linenosize=\small} \algsetup{indent=1em}
    \begin{algorithmic}[1]
    \STATE{$\mathcal{P}_{l} = \emptyset$ for $l = 0, ..., l_f$};\\
     \STATE{[${\mathcal{G}_1},...,{\mathcal{G}_{l_f}};\mathbb{H}_1,...,\mathbb{H}_{l_f}$] = Multilevel$\_$spectral$\_$graph$\_$reduction (${\mathcal{G}}$)};\\
     \STATE{${\mathcal{P}_{l_f}}={\mathcal{G}_{l_f}}$; $l=l_f$}\\
     
    \WHILE{ $ l\ge1$}
        \FOR{each node $i \in \mathcal{V}_l$}
         %   \STATE{Form graph $\mathcal{F}_{l-1}^{(i)}=(\tilde{\mathcal{V}}_{l-1}^{(i)},\tilde{\mathcal{E}}_{l-1}^{(i)}, \omega)$ with $\tilde{\mathcal{V}}_{l-1}^{(i)} = \mathcal{S}_{l-1}^{(i)}$ and $\tilde{\mathcal{E}}_{l-1}^{(i)} \subset \mathcal{E}_{l-1}.$}  \COMMENT{Find the subset graph  $\mathcal{F}_{l-1}^{(i)}$ of $\mathcal{G}_{l-1}$ formed by the nodes in set $\mathcal{S}_{l-1}^{(i)}$} ;
         \STATE{Find the induced subgraph  $\mathcal{F}_{l-1}^{(i)}$  formed by the nodes set $\mathcal{S}_{l-1}^{(i)}$} in $\mathcal{G}_{l-1}$ ;
            \STATE{Extract the LSST $\mathcal{T}_{l-1}^{(i)}$ of $\mathcal{F}_{l-1}^{(i)}$ };
    
            \STATE{$\mathcal{P}_{l-1} = \mathcal{P}_{l-1} \cup \mathcal{T}_{l-1}^{(i)}$};
        \ENDFOR %\vspace{0.1in}
        \FOR{each edge ${(p,q)} \in \mathcal{}{E}_l$}
            \STATE{ Node set $s_p = S_{l-1}^{(p)}$, node set $s_q = S_{l-1}^{(q)}$}; \\
            Find the edge with maximum weight between $s_p$ and $s_q$, and add the edge into $\mathcal{P}_{l-1}$ ;
           % \STATE{find maximum edge between $s_p$ and $s_q$, and add the edge into $\mathcal{P}_{l-1}$ };
    
           % \STATE{$\mathcal{P}_{l-1} = \mathcal{P}_{l-1} \cup \mathcal{T}_{l-1}^i$};
        \ENDFOR 
        
        \STATE{Embed $\mathcal{G}_{l-1}$ to $K$-dimensional space $X_{l-1}=[x_{l-1}^{(1)}, x_{l-1}^{(2)},...,x_{l-1}^{(K)}]$};
        %\STATE{For each subgraph edge $(p,q) \in (\mathcal{E}_{l-1}-E_{l-1})$, calculate the edge distortion $d(p,q) = {\omega_{l-1}(p,q)} \|X_{l-1}(p, :)-X_{l-1}(q,:)\|$};
        \STATE{For each subgraph edge $(p,q) \in (\mathcal{E}_{l-1}-E_{l-1})$, calculate the edge distortion $d(p,q) \propto {\omega_{l-1}(p,q)} \left( X^\top_{l-1} e_{p,q} \right)^2$};
        \STATE{Include top few edges with large distortion into $\mathcal{P}_{l-1}$};
    \STATE{$l=l-1$}
   % \STATE{$\textrm{iter}=\textrm{iter}+1$};
    \ENDWHILE\\
     \STATE{let ${\mathcal{P}}={\mathcal{P}_{0}}$ and return graph ${\mathcal{P}}$};
    
    \end{algorithmic}
    }
    
\end{algorithm} 
 Algorithm \ref{alg:directed_graph_spar} shows the algorithm flow for the proposed SF-GRASS framework. The complexity of spectral graph coarsening is $O(|\mathcal{E}_l|)$ for each level, the complexity of backward graph mapping procedure is $O(|\mathcal{E}_l|)$ for each level $l$, and the complexity of off-subgraph identification is $O(|\mathcal{E}_l|)$ for each level $l$. If the spectral coarsening step will produce $O(\log |\mathcal{V}|)$  graphs with a fixed coarsening ratio for two consecutive levels, the overall runtime complexity of SF-GRASS is nearly linear for an input graph $\mathcal{G}=(\mathcal{V},\mathcal{E}, \omega)$.
%  \vspace{-0.3cm}

\section{Experimental results}\label{result_sec}

The proposed spectral sparsification algorithm has been implemented in Matlab and $C$++. 
The test cases used in this paper have been selected from a great variety of matrices that have been used in circuit simulation, finite element analysis, machine learning, and data mining applications. If the original matrix is not a graph Laplacian, it will be converted into a graph Laplacian by setting each edge weight using the absolute value of each nonzero entry in the lower triangular matrix; if edge weights are not available in the original matrix file, a unit edge weight will be assigned to all edges. All of our experiments have been conducted using a single CPU core of a computing platform running 64-bit RHEW 7.2 with a $2.67$GHz 12-core CPU and $50$ GB memory. Several test cases have been tested in the experiments.

\subsection{SF-GRASS for Spectral Graph Sparsification}\label{result:sf_grass}

 \begin{table*}
 \scriptsize
 \centering \caption{Comparison of spectral sparsification results between SF-GRASS and GRASS. }
 \begin{tabular}{|c|c|c|c|c|c|c|c|c|c|c|c|}
  \hline
 \multirow{2}{*}{Test cases} & {\multirow{2}{*}{$\mathcal{N}$}} & {\multirow{2}{*}{$\mathcal{M}$}} &  \multicolumn{3}{|c|}{GRASS} & \multicolumn{6}{|c|}{SF-GRASS}\\
\cline{4-12}
 { } & { } & { } & {$T_{grass}$} & {$\frac{|\mathcal{E}_{off}|}{\mathcal{N}}$} & {$\kappa(\mathcal{L}_{\mathcal{G}},\mathcal{L}_{\mathcal{P}})$} & {$T_{r}$} & {$T_{spar}$} & {$\frac{|\mathcal{E}_{off}|}{\mathcal{N}}$} & {$\kappa(\mathcal{L}_{\mathcal{G}},\mathcal{L}_{\mathcal{P}})$} & {$\kappa(\mathcal{L}_{\mathcal{G}},\mathcal{L}_{\mathcal{S}})$} & {$\frac{\kappa(\mathcal{L}_{\mathcal{G}},\mathcal{L}_{\mathcal{S}})}{\kappa(\mathcal{L}_{\mathcal{G}},\mathcal{L}_{\mathcal{P}})}$}\\
  \hline
  {fe$\_$4elt} & {$1.1E4$} & {$3.3E4$} & {$0.10s$} & {$21.6\%$} & {$50$} & {$0.01s $} & {$0.16s$} & {$19.3\%$} & {$51$} & {$6.36E4$ } & {$1.25E3X$ }\\
  \hline
  
  {fe$\_$ocean} & {$1.4E5$} & {$4.1E5$} & {$2.21s$} & {$9.0\%$} & {$271$} & {$1.37s $} & {$0.17s$} & {$9.7\%$} & {$276$} & { $2.16E6$} & {$7.82E3X$}\\
  \hline
  
%  {vsp$\_$befref$\_$fxm} & {$1.4E4$} & {$9.8E4$} & {$0.13s$} & {$2.6\%$} & {$7419$} & {$ $} & {$0.15s$} & {$2.6\%$} & {$5545$}\\
  %\hline
  
  {Gmat$\_$airfoil} & {$4.3E4$} & {$1.2E4$} & {$0.03s$} & {$7.4\%$} & {$131$} & {$ 0.08s$} & {$0.06s$} & {$7.5\%$} & {$99$} & {$1.10E4$ } & {$1.11E2X$}\\
  \hline
  
  {G2$\_$circuit} & {$1.5E5$} & {$2.9E5$} & {$1.24s$} & {$3.0\%$} & {$306$} & {$1.26s $} & {$0.12s$} & {$3.1\%$} & {$423$} & {$1.43E5$  } & {$3.39E2X$ }\\
  %  \hline
%{G2$\_$circuitl} & {$1.5E5$} & {$2.9E5$} & {$1.19s$} & {$1.5\%$} & {$1108$} & {$ $} & {$0.61s$} & {$1.5\%$} & {$1037$}\\
    \hline
    
  {Gmat$\_$laplacian$\_0.25M$} & {$2.5E5$} & {$7.4E5$} & {$4.89s$} & {$10.0\%$} & {$238$} & {$2.47s $} & {$0.26s$} & {$10.0\%$} & {$357$} & { $6.31E6$} & {$1.77E4X$ }\\
   \hline
   
  %{Gmat$\_$ibm3} & {$8.5E5$} & {$1.4E6$} & {$10.15s$} & {$2.2\%$} & {$89$} & {$ 7.01s$} & {$0.70s$} & {$3.7\%$} & {$85$} & { } & { }\\
   %\hline
    %{Gmat$\_$ibm5} & {$1.1E6$} & {$1.6E6$} & {$9.64s$} & {$2.1\%$} & {$87$} & {$8.49s $} & {$0.74s$} & {$2.3\%$} & {$144$} & { } & { }\\
   
  %{Gmat$\_$ibm5} & {$1.1E6$} & {$1.6E6$} & {$32.21s$} & {$2.1\%$} & {$87$} & {$ $} & {$1.32s$} & {$2.3\%$} & {$144$}\\

   %\hline
 %{PPI} & {$3.9E3$} & {$3.8E4$} & {$3.3\%$} & {$0.04s$} & {$4685$} & {$0.24s$} & {$3495$}\\
  %\hline
 %{fe$\_$4elt} & {$1.1E4$} & {$3.3E4$} & {$1.4\%$} & {$0.08s$} & {$472$} & {$0.25s$} & {$451$}\\
 %\hline
 %{Gmat$\_$airfoil} & {$4.3E4$} & {$1.2E4$} & {$2.4\%$} & {$0.03s$} & {$477$} & {$0.11s$} & {$422$}\\
 %\hline
  %{vsp$\_$befref$\_$fxm} & {$1.4E4$} & {$9.8E4$} & {$2.6\%$} & {$0.08s$} & {$4483$} & {$0.68s$} & {$2547$}\\
  \hline\end{tabular}\label{table:sf_grass}
 \end{table*}
 
Table \ref{table:sf_grass} shows the spectral graph sparsification results on various graphs when comparing to the state-of-the-art sparsification tool GRASS\footnote{ \url{https://sites.google.com/mtu.edu/zhuofeng-graphspar/home}} \cite{feng2016spectral, zhuo:dac18,feng2020grass}, where $\mathcal{N}$ ($\mathcal{M}$) represents the number of nodes (edges) in the original graph; $T_{grass}$ denotes the the sparsifier construction time using GRASS; $T_r$ denotes the multilevel graph coarsening time; $T_{spar}$ denotes the multilevel sparsifier construction time by SF-GRASS; $|\mathcal{E}_{off}|$ denotes the number of off-tree edges added  for forming the final sparsifier from the initial spanning-tree sparsifier. $\kappa(\mathcal{L}_{\mathcal{G}}, \mathcal{L}_{\mathcal{P}})$ denotes the final relative condition number between the Laplacians of the original graph $\mathcal{G}$ and the sparsifier $\mathcal{P}$.  $\kappa(\mathcal{L}_{\mathcal{G}}, \mathcal{L}_{\mathcal{S}})$ denotes the relative condition number between the Laplacians of the original graph $\mathcal{G}$ and the initial spanning-tree sparsifier $\mathcal{S}$ generated by SF-GRASS. When the original graph is relatively small, the runtime of GRASS and SF-GRASS are comparable. However,   SF-GRASS can become substantially faster  when confronting greater graph sizes and densities since GRASS requires a Laplacian solver to compute dominant generalized eigenvectors while SF-GRASS does not.

\begin{figure*}[!htb]
\minipage{0.43409\textwidth}
  \includegraphics[width=\linewidth]{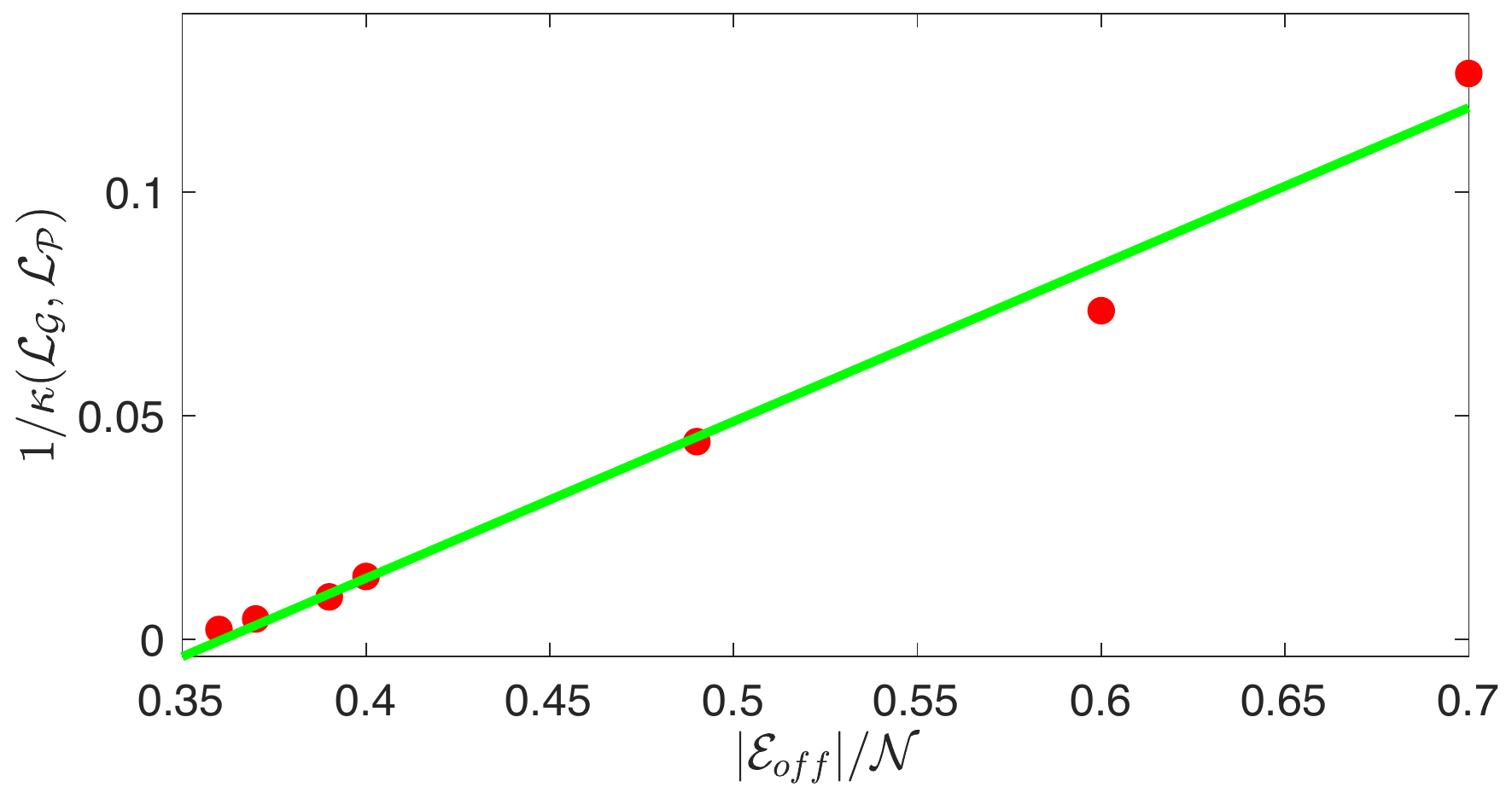}
  \caption{Condition number change with number of off-tree edges added for fe$\_$4elt graph}\label{fig:fe_4felt_conditionnumber_vs_added off-tree}
\endminipage\hfill
\minipage{0.43409\textwidth}
  \includegraphics[width=\linewidth]{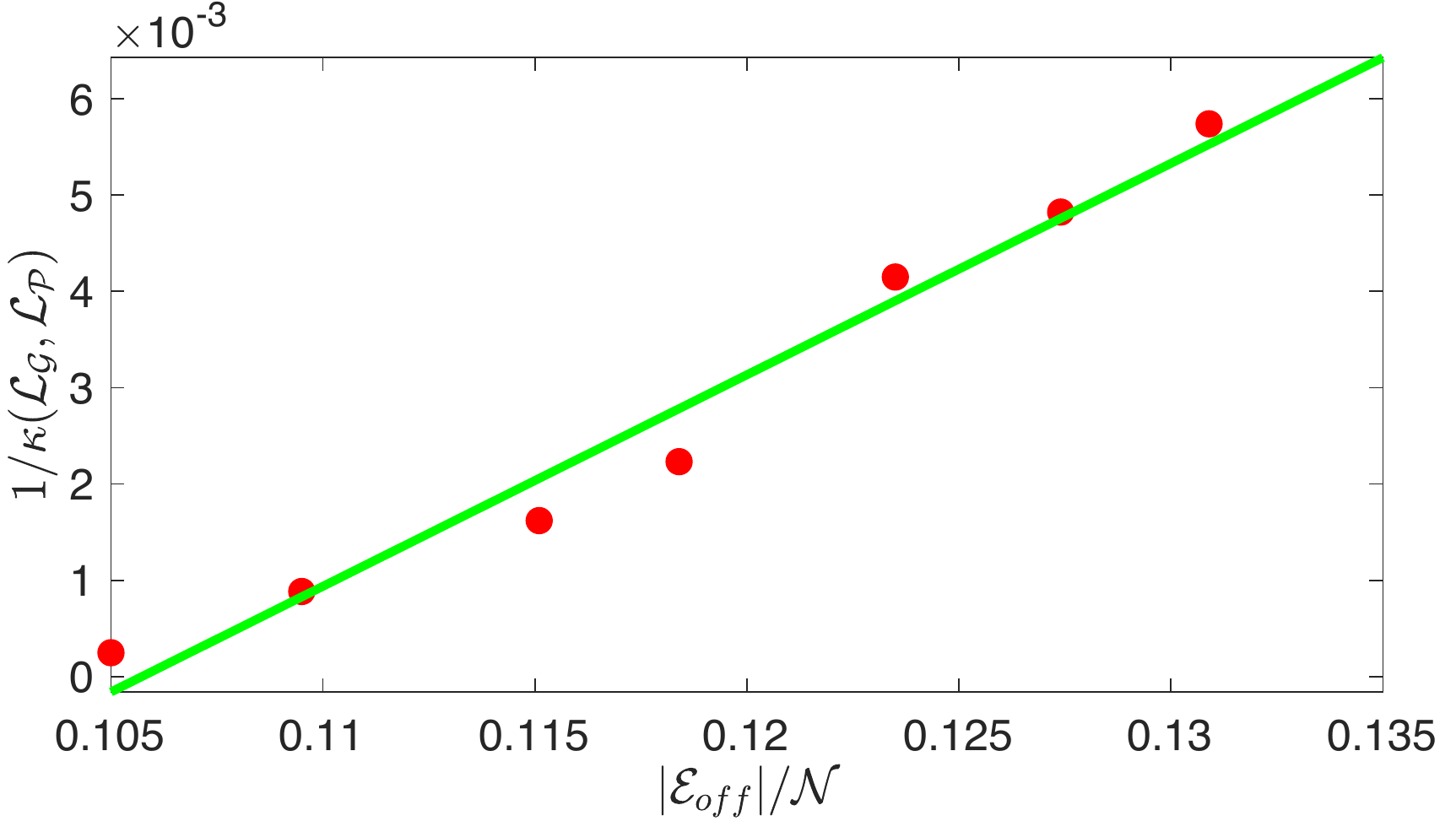}
  \caption{Condition number change with number of off-tree edges added for PPI graph}\label{fig:ppi_conditionnumber_vs_added off-tree}
\endminipage
\end{figure*}

Figure \ref{fig:fe_4felt_conditionnumber_vs_added off-tree} and Figure \ref{fig:ppi_conditionnumber_vs_added off-tree} show the changes of the relative condition numbers with increasing number of  off-tree edges added to the initial spanning-tree sparsifier for  PPI and fe$\_$4elt graphs. It can be observed that smaller condition number can be achieved with greater number of off-tree edges included, which indicates that very desired (flexible) tradeoffs between graph complexity and approximation quality can be obtained.

\subsection{SF-GRASS for PCG Iterations}\label{result:PCG}
\begin{table*}
 \scriptsize
 \centering \caption{Results of the PCG solver for SF-GRASS. }
 \begin{tabular}{|c|c|c|c|c|c|c|c|c|c|c|c|}
  \hline
 \multirow{2}{*}{Test cases} & {\multirow{2}{*}{$\mathcal{N}$}} & {\multirow{2}{*}{$\mathcal{M}$}} & {\multirow{2}{*}{directed solver}} & \multicolumn{4}{|c|}{PCG for SF-GRASS} & \multicolumn{4}{|c|}{PCG for GRASS}\\
\cline{5-12}
 { } & { } & { } & { } & {$T$} &{$\frac{|\mathcal{E}_{off}|}{\mathcal{N}}$}& {$iter$} & {$relres$} & {$T$}&{$\frac{|\mathcal{E}_{off}|}{\mathcal{N}}$} & {$iter$} & {$relres$}\\
  \hline
{Thermal1} & {$2.5E4$} & {$7.2E4$} & {$1.12s$} & {$0.13s$} & {$2.7\%$} &{$3$}&{$5.6E-4$}& {$0.26s$} & {$2.6\%$} & {$7$}&{$4.5E-4$}\\
  \hline
{Thermal2} & {$1.0E5$} & {$2.9E5$} & {$5.95s$} & {$0.79s$} & {$2.9\%$} &{$3$}&{$5.9E-4$}& {$3.31s$} & {$2.9\%$} & {$7$}&{$5.7E-4$}\\
  \hline
  {Thermal3} & {$2.0E5$} & {$5.9E5$} & {$19.42s$} & {$3.41s$} & {$3.2\%$} &{$4$}&{$1.8E-4$}& {$9.44s$} & {$2.9\%$} & {$8$}&{$6.3E-4$}\\
  \hline
    {Thermal4} & {$4.0E5$} & {$1.2E6$} & {$72.47s$} & {$8.01s$} & {$3.1\%$} &{$3$}&{$6.0E-4$}& {$63.52s$} & {$2.9\%$} & {$6$}&{$8.6E-4$}\\
  \hline
    {Thermal5} & {$9.0E5$} & {$2.6E6$} & {$974.87s$} & {$21.28s$} & {$3.1\%$} &{$3$}&{$6.2E-4$}& {$919.35s$} & {$3.0\%$} & {$6$}&{$8.2E-4$}\\
  \hline
   {Thermal6} & {$1.6E6$} & {$4.6E6$} & {$3637.79s$} & {$42.02s$} & {$3.1\%$} &{$3$}&{$6.1E-4$}& {$1695.92s$} & {$3.0\%$} & {$6$}&{$5.6E-4$}\\
   \hline
   {Thermal7} & {$2.0E6$} & {$5.9E6$} & {$2787.94s$} & {$69.62s$} & {$3.2\%$} &{$3$}&{$6.2E-4$}& {$7932.55s$} & {$3.0\%$} & {$6$}&{$6.2E-4$}\\
      \hline
   {Thermal8} & {$2.5E6$} & {$7.2E6$} & {$9341.48s$} & {$56.36s$} & {$3.2\%$} &{$3$}&{$6.1E-4$}& {$5476.88s$} & {$3.0\%$} & {$6$}&{$6.2E-4$}\\
  
  \hline\end{tabular}\label{tab_total}
 \end{table*}

 The spectral sparsifier generated by the proposed algorithm is leveraged as a preconditioner in a PCG solver for solving linear system equations $Ax = b$. The preconditioner matrix is factorized with Cholmod solver \cite{cholmod}. The right-hand-side (RHS) vector $b$ is generated randomly, while the solver is set to converge to an accuracy level $\|Ax-b\|/\|b\|<1E-3$ for all test cases. We compare the PCG solver using SF-GRASS with the direct method and the PCG solver using GRASS, as shown in Table \ref{tab_total}. $T$ represents the total runtime for each solver, $iter$ represents the number of iterations, and $relres$ is the relative residue. It shows that SF-GRASS is the fastest among all three solvers, which can achieve up to $167X$ and $98X$ speedups when comparing to direct solver and PCG solver using GRASS, respectively. Also, SF-GRASS has achieved a faster convergence rate than GRASS. 
 
  \begin{figure}[!htb]
  \includegraphics[width=0.95\linewidth]{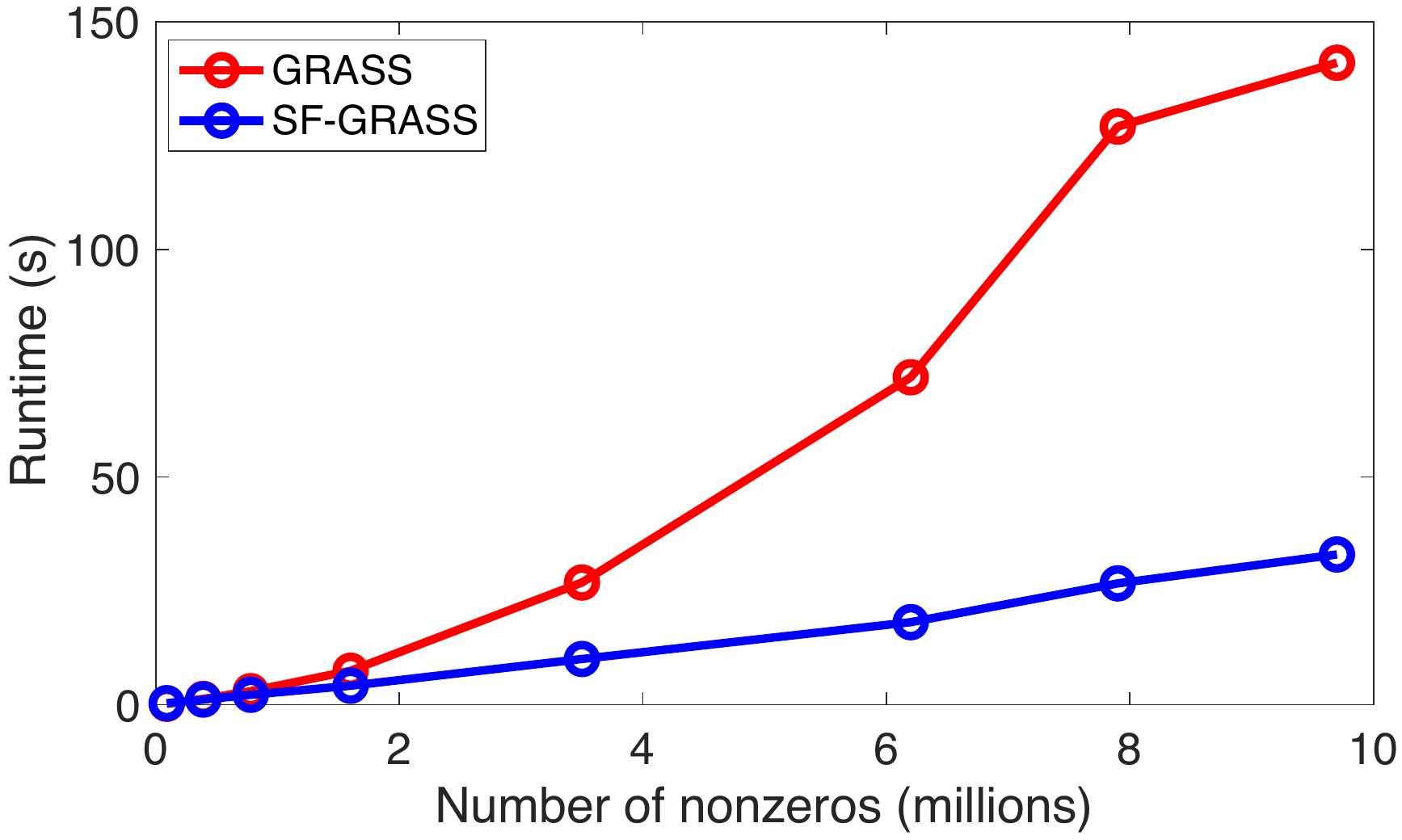}
  \caption{Runtime scalability comparison: GRASS vs SF-GRASS  (3D meshes of different sizes) }\label{fig:scalability}
  \end{figure}
  
  Figure \ref{fig:scalability} shows the runtime scalability of GRASS and SF-GRASS on different sizes of 3D mesh graphs. It indicates that  SF-GRASS scales linearly with the graph size, which is more scalable than GRASS,  especially on larger and denser graphs, such as 3D mesh graphs.
  
  \begin{figure}[!htb]
 \includegraphics[width=0.95\linewidth]{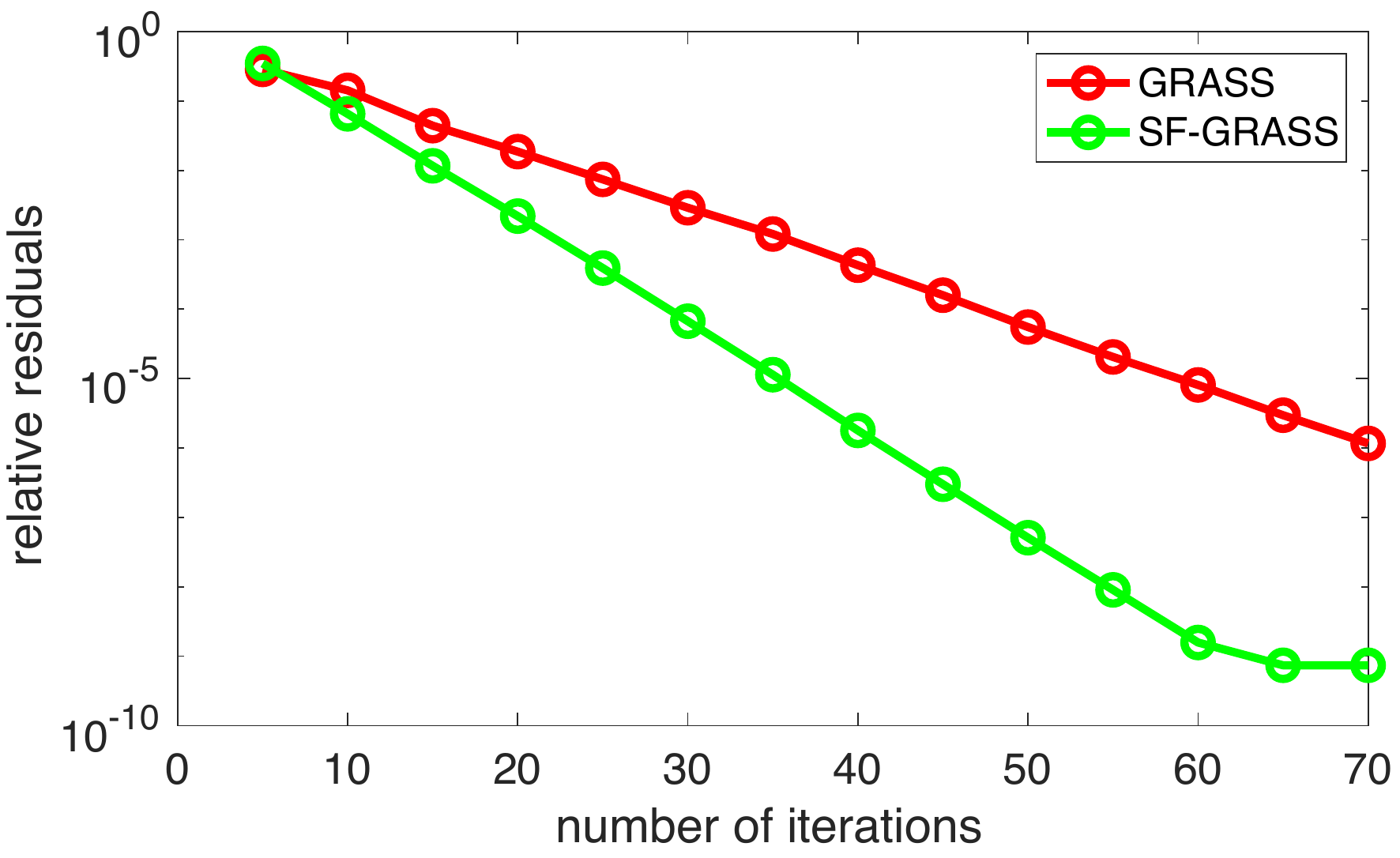}
 %\caption{Convergence rate for mat$\_$0.1M comparison: ichol, GRASS vs SF-GRASS 
 \caption{Convergence rate comparison for a 3D thermal grid:  GRASS vs SF-GRASS
 }\label{fig:convergency}
\end{figure}

Figure \ref{fig:convergency} shows the convergence rate of PCG solver when using the sparsifiers generated by GRASS and SF-GRASS on a 3D thermal grid with $1.0E5$   nodes and $3.0E5$   edges. To generate the sparsifiers, we add ${0.018\mathcal{N}}$ off-tree edges to the sparsifiers for both SF-GRASS and GRASS settings. It shows that SF-GRASS has achieved a better convergence rate than GRASS.  % Also, we perform incomplete Cholesky factorization (ichol) to get the third preconditioner. Then we do the  preconditioned conjugate gradients method (pcg) to get the relative residuals at different number of iteration to between Laplacian matrix and its three sparsifier. We can see the convergence rates in Fig  \ref{fig:convergency} and we can claim that SF-GRASS converges faster than both GASS and ichol. 

\begin{figure}[!htb]
  \includegraphics[width=0.95\linewidth]{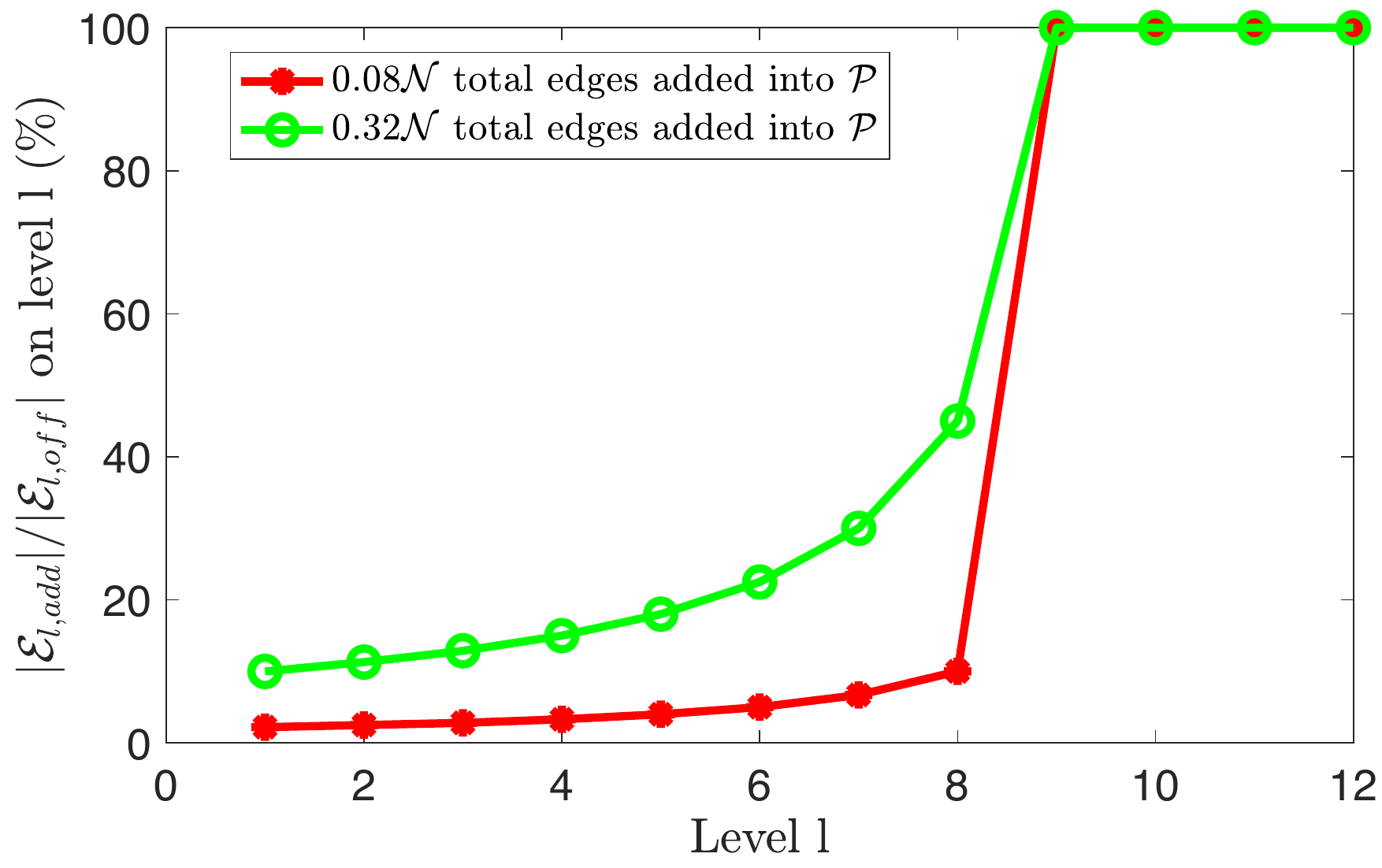}
  \caption{Edge sampling probabilities for each coarse-level graph of Thermal3, where total $0.08\mathcal{N}$ and  $0.32\mathcal{N}$ number of edges are added into $\mathcal{P}$, respectively }\label{fig:probability}
\end{figure}

% \begin{figure}[!htb]
%   \includegraphics[width=0.98\linewidth]{bar_eff.pdf}
%   \caption{Edge sampling probabilities for each coarse-level graph of Thermal3, where total $0.08\mathcal{N}$ and  $0.32\mathcal{N}$ number of edges are added into $\mathcal{P}$, respectively }\label{fig:bar_eff}.
% \end{figure}

% \begin{figure}[!htb]
%   \includegraphics[width=0.98\linewidth]{mesh.pdf}
%   \caption{average effective resistance for mesh}\label{fig:mesh_eff}.
% \end{figure}

% \begin{figure}[!htb]
%   \includegraphics[width=0.98\linewidth]{airfoil.pdf}
%   \caption{average effective resistance for airfoil}\label{fig:airfoil_eff}.
% \end{figure}

% \begin{figure}[!htb]
%   \includegraphics[width=0.98\linewidth]{delaunay.pdf}
%   \caption{average effective resistance for denaulay10 }\label{fig:denaulay_eff}.
% \end{figure}

Figure \ref{fig:probability} shows the edge sampling probabilities on each coarse level   graph of a 3D thermal mesh grid, where $0.08\mathcal{N}$ and  $0.32\mathcal{N}$  off-tree edges have been added to the initial spanning-tree sparsifier $\mathcal{P}$ across all levels, respectively. $|\mathcal{E}_{l,add}|$ denotes the number of off-subgraph edges added on level $l$ graph, and $|\mathcal{E}_{l,off}|$ denotes the total off-subgraph edges on level $l$. As shown, the edges on coarser graphs  have been assigned with higher sampling probabilities since they will have greater effective-resistances and thus be more important for retaining the original graph  structural (spectral) properties.

 %\vspace*{-10pt}
 \subsection{SF-GRASS for Vectorless   Verification}\label{result:verification}
 \begin{table*}
\scriptsize
\centering \caption{Results of the proposed vectorless power grid integrity verification method. }
\begin{tabular}{|c|c|c|c|c|c|c|c|c|c|c|c|c|c|c|c|}
 \hline
  {}& \multicolumn{3}{|c|}{Power Grid Specs.}& \multicolumn{3}{|c|}{Single Level} & \multicolumn{4}{|c|}{Multilevel w/o Sparsifier} & \multicolumn{5}{|c|}{Multilevel w/ Sparsifer}\\
 \hline $CKT$ &{$\mathcal{N}.\#$}&{$\mathcal{C}.\#$} &{${L}. \#$}&{$T_{chol}$}& $T_{sol}$ & $T_{lp}$ & $T_{chol}$  & $T_{sol}$ & $T_{lp}$ & $Err(\%)$& $T_{chol}$ & $T_{sol}$ & $T_{lp}$ & $Err(\%)$ & {$\kappa$}\\

 \hline $ibmpg3$ &{$8.5E5$}&{$9.0E4$} &{$2$}&{$11.91s$}  & $0.40 s$ & $1.63 s$ & $15.55 s$ & $0.51 s$ & $0.05 s$ & $1.76\%$& $1.70 s$ & $0.04 s$ & $0.03 s$ & $1.85\%$ & $239$   \\ % , 1.9\%
 \hline $ibmpg4$ &{$1.0E6$}&{$1.0E5$} &{$2$}&{$14.97 s$} & $0.53 s$ & $1.67 s$ & $20.99 s$  & $0.73 s$ & $0.14 s$ & $2.71\%$& $1.68 s$ & $0.04 s$ & $0.10 s$ & $3.52\%$ & $1136$  \\ % , 1.1\%
 \hline $ibmpg5$ &{$1.1E6$}&{$1.6E5$} &{$2$}&{$8.48 s$} & $0.27 s$ & $2.08 s$ & $12.58 s$  & $0.43 s$ & $0.22 s$ & $2.43\%$& $2.10 s$ & $0.05 s$ & $0.17 s$ & $2.52\%$ & $218$  \\  %, 1.8\%
 \hline $ibmpg6$ &{$1.7E6$}&{$1.7E5$} &{$2$}&{$12.24s$}  & $0.36 s$ & $3.21 s$  & $17.76 s$ & $0.51 s$ & $0.20 s$& $1.36\%$ & $3.05 s$ & $0.07 s$ & $0.06 s$ & $3.83\%$ & $248$ \\ %, 1.3\%
 %\hline $ibmpg7$ &{$1.5E6$}&{$1.5E5$} &{$2$} & $22.79 s$ & $0.71 s$ & $2.83 s$  & $30.47 s$ & $0.97 s$ & $0.10 s$& $3.32\%$  & $4.86 s$ & $0.06 s$ & $0.15 s$& $7.31\%$ & $382, 2.2\%$    \\
 \hline $thupg1$ &{$5.0E6$}&{$5.0E5$} &{$2$}&{$72.44 s$}  & $2.07 s$ & $9.40 s$ & $290.36 s$  & $4.06 s$ & $28.31 s$ & $1.73\%$& $11.96 s$ & $0.25 s$ & $4.93 s$ & $3.31\%$ & $464$     \\ % ,2.17\%
 \hline $thupg2$ &{$9.0E6$}&{$9.0E5$} &{$2$}&{$955.00s$}  & $4.53 s$ & $33.40 s$ & $1142.46 s$  & $6.59 s$ & $14.26 s$ & $4.20\%$ & $52.75 s$ & $0.50 s$ & $9.90 s$ & $2.64\%$ & $465$  \\ %2.17
 %\hline $CKT7$ &$1.02M$&{$300K$} &{$2$}&{$1600$}  & $2, 981 s$ & $1,070 s$ & $4,051 s$  & $1,932 s$ & $400 s$ & $2,332 s$& $6~mV$ & $1,110 s$ & $306 s$ & $1,416 s$    \\
 \hline\end{tabular}\label{table:ver}
\end{table*}
We also evaluated SF-GRASS for vectorless power grid verifications  using industrial power gird designs with different sizes \cite{ibmpg}, as shown in Table \ref{table:ver}. The vectorless verification framework is adopted from \cite{zhiqiang:dac17}. ``Single Level", ``Multilevel w/o Sparsifier", and ``Multilevel w/ Sparsifier"  denote the  verification methods using single level (direct), multilevel grids w/o sparsification and w/ sparsification using SF-GRASS, respectively. Note that we choose to apply sparsified power grid on each level generated by SF-GRASS during the verification process. $\mathcal{N}.\#$, $\mathcal{C}.\#$, ${L}.\#$ are the numbers of grid nodes, current sources, and hierarchical levels, respectively. $T_{chol}$, $T_{sol}$ and $T_{lp}$ denote the runtime for Cholesky factorizations, adjoint sensitivity calculation using matrix factors and the total LP solution time including all levels, respectively. $Err$ denotes the relative error of maximum voltage drop compared to the single-level method, and $\kappa$ denotes the relative condition number.

 For all test cases, it is observed that matrix factorization, sensitivity calculation, and LP solving can be significantly accelerated using the SF-GRASS while maintaining excellent accuracy. The ``Multilevel w/o Sparsifier" method is always the slowest due to the fast-growing matrix densities at coarse levels.

%\begin{figure*}[!htb]
%\minipage{0.45\textwidth}
%  \includegraphics[height=3.5cm,width=\linewidth]{probability_drawing_8.pdf}
 % \caption{Edge sampling probabilities for various coarse-level graphs (Thermal3) with  adding 8 $\%$  $\mathcal{N}$ edges into $\mathcal{P}$ }\label{fig:probability 8}
%\endminipage\hfill
%\minipage{0.45\textwidth}
%  \includegraphics[height=3.5cm,width=\linewidth]{probability_drawing.pdf}
 % \caption{Edge sampling probabilities for various coarse-level graphs (Thermal3) with adding 32 $\%\mathcal{N}$ edges into $\mathcal{P}$  }\label{fig:probability32}
%\endminipage
%\end{figure*}

\section{Conclusions}\label{conclusion}
For the first time, we present a solver-free spectral graph sparsification approach (SF-GRASS)   by leveraging emerging spectral graph coarsening and  graph signal processing (GSP) techniques.  Such a scalable framework allows constructing a hierarchy of spectrally-reduced and sparsified graphs in nearly-linear time, which can become key to accelerating many graph-based numerical computing tasks. The proposed spectral approach is simple to implement and inherently parallel friendly. Our extensive experimental results show that the proposed method can produce a hierarchy of high-quality spectral  sparsifiers in nearly-linear time  for a variety of real-world, large-scale graphs and circuit networks when compared with prior state-of-the-art spectral methods.
% \section{Acknowledgments}
% This work is supported in part by  the National Science Foundation under Grants  CCF-1350206 and CCF-1618364.
%  \vspace{-0.03cm}
% \bibliographystyle{unsrt}
%  \vspace{-0.25cm}

 \section{Acknowledgments}
This work is supported in part by  the National Science Foundation under Grants  CCF-1350206 (CAREER),   CCF-2021309 (SHF), and CCF-2011412 (SHF).
%  \vspace{-0.25cm}
\bibliographystyle{abbrv}
{
\bibliography{dac20_spectral,nsf19,feng}  % sigproc.bib is the name of the Bibliography in this case

\begin{thebibliography}{10}

\bibitem{abraham2012}
I.~Abraham and O.~Neiman.
\newblock Using petal-decompositions to build a low stretch spanning tree.
\newblock In {\em Proceedings of the forty-fourth annual ACM symposium on
  Theory of computing (STOC)}, pages 395--406. ACM, 2012.

\bibitem{batson2012twice}
J.~Batson, D.~Spielman, and N.~Srivastava.
\newblock {Twice-Ramanujan Sparsifiers}.
\newblock {\em SIAM Journal on Computing}, 41(6):1704--1721, 2012.

\bibitem{benczur1996approximating}
A.~A. Bencz{\'u}r and D.~R. Karger.
\newblock Approximating st minimum cuts in {\~o} (n 2) time.
\newblock In {\em Proceedings of the twenty-eighth annual ACM symposium on
  Theory of computing (STOC)}, pages 47--55. ACM, 1996.

\bibitem{benczur2015randomized}
A.~A. Bencz{\'u}r and D.~R. Karger.
\newblock Randomized approximation schemes for cuts and flows in capacitated
  graphs.
\newblock {\em SIAM Journal on Computing}, 44(2):290--319, 2015.

\bibitem{cholmod}
T.~Davis.
\newblock {\em CHOLMOD: sparse supernodal Cholesky factorization and
  update/downdate}.
\newblock [Online]. Available:
  http://www.cise.ufl.edu/research/sparse/cholmod/, 2008.

\bibitem{deng2019graphzoom}
C.~Deng, Z.~Zhao, Y.~Wang, Z.~Zhang, and Z.~Feng.
\newblock Graphzoom: A multi-level spectral approach for accurate and scalable
  graph embedding.
\newblock {\em International Conference on Learning Representations (ICLR)},
  2020.

\bibitem{elkin2008lower}
M.~Elkin, Y.~Emek, D.~A. Spielman, and S.-H. Teng.
\newblock Lower-stretch spanning trees.
\newblock {\em SIAM Journal on Computing}, 38(2):608--628, 2008.

\bibitem{feng2016spectral}
Z.~Feng.
\newblock Spectral graph sparsification in nearly-linear time leveraging
  efficient spectral perturbation analysis.
\newblock In {\em Proceedings of the 53rd Annual Design Automation Conference},
  pages 1--6, 2016.

\bibitem{zhuo:dac18}
Z.~Feng.
\newblock Similarity-aware spectral sparsification by edge filtering.
\newblock In {\em Design Automation Conference (DAC), 2018 55nd ACM/EDAC/IEEE},
  pages 1--6. IEEE, 2018.

\bibitem{feng2020grass}
Z.~Feng.
\newblock Grass: Graph spectral sparsification leveraging scalable spectral
  perturbation analysis.
\newblock {\em IEEE Transactions on Computer-Aided Design of Integrated
  Circuits and Systems}, 2020.

\bibitem{lengfei:tcad15}
L.~Han, X.~Zhao, and Z.~Feng.
\newblock {An Adaptive Graph Sparsification Approach to Scalable Harmonic
  Balance Analysis of Strongly Nonlinear Post-Layout RF Circuits}.
\newblock {\em Computer-Aided Design of Integrated Circuits and Systems, IEEE
  Transactions on}, 34(2):173--185, 2015.

\bibitem{hong2018efficient}
C.~Hong, A.~Sukumaran-Rajam, B.~Bandyopadhyay, J.~Kim, S.~E. Kurt, I.~Nisa,
  S.~Sabhlok, {\"U}.~V. {\c{C}}ataly{\"u}rek, S.~Parthasarathy, and
  P.~Sadayappan.
\newblock Efficient sparse-matrix multi-vector product on gpus.
\newblock In {\em Proceedings of the 27th International Symposium on
  High-Performance Parallel and Distributed Computing}, pages 66--79. ACM,
  2018.

\bibitem{kelner2014almost}
J.~A. Kelner, Y.~T. Lee, L.~Orecchia, and A.~Sidford.
\newblock {An Almost-linear-time Algorithm for Approximate Max Flow in
  Undirected Graphs, and Its Multicommodity Generalizations}.
\newblock In {\em Proceedings of the twenty-fifth annual ACM-SIAM symposium on
  Discrete algorithms}, pages 217--226. SIAM, 2014.

\bibitem{kipf2016semi}
T.~N. Kipf and M.~Welling.
\newblock Semi-supervised classification with graph convolutional networks.
\newblock {\em arXiv e-print, arXiv:1609.02907}, 2016.

\bibitem{lee2014multiway}
J.~R. Lee, S.~O. Gharan, and L.~Trevisan.
\newblock {Multiway spectral partitioning and higher-order cheeger
  inequalities}.
\newblock {\em Journal of the ACM (JACM)}, 61(6):37, 2014.

\bibitem{Lee:2017}
Y.~T. Lee and H.~Sun.
\newblock {An SDP-based Algorithm for Linear-sized Spectral Sparsification}.
\newblock In {\em Proceedings of the 49th Annual ACM SIGACT Symposium on Theory
  of Computing}, STOC 2017, pages 678--687, New York, NY, USA, 2017. ACM.

\bibitem{livne2012lean}
O.~Livne and A.~Brandt.
\newblock {Lean algebraic multigrid (LAMG): Fast graph Laplacian linear
  solver}.
\newblock {\em SIAM Journal on Scientific Computing}, 34(4):B499--B522, 2012.

\bibitem{loukas2019graph}
A.~Loukas.
\newblock Graph reduction with spectral and cut guarantees.
\newblock {\em Journal of Machine Learning Research}, 20(116):1--42, 2019.

\bibitem{loukas2018spectrally}
A.~Loukas and P.~Vandergheynst.
\newblock Spectrally approximating large graphs with smaller graphs.
\newblock In {\em International Conference on Machine Learning}, pages
  3243--3252, 2018.

\bibitem{ibmpg}
S.~R. Nassif.
\newblock {\em {IBM power grid benchmarks}}.
\newblock [Online]. Available: http://dropzone.tamu.edu/~pli/PGBench/, 2008.

\bibitem{peng2013phd}
R.~Peng.
\newblock {\em Algorithm Design Using Spectral Graph Theory}.
\newblock PhD thesis, Carnegie Mellon University, 2013.

\bibitem{peng2015partitioning}
R.~Peng, H.~Sun, and L.~Zanetti.
\newblock Partitioning well-clustered graphs: Spectral clustering works.
\newblock In {\em Proceedings of The 28th Conference on Learning Theory
  (COLT)}, pages 1423--1455, 2015.

\bibitem{shuman2013emerging}
D.~I. Shuman, S.~K. Narang, P.~Frossard, A.~Ortega, and P.~Vandergheynst.
\newblock The emerging field of signal processing on graphs: Extending
  high-dimensional data analysis to networks and other irregular domains.
\newblock {\em IEEE Signal Processing Magazine}, 30(3):83--98, 2013.

\bibitem{spielman2011graph}
D.~Spielman and N.~Srivastava.
\newblock Graph sparsification by effective resistances.
\newblock {\em SIAM Journal on Computing}, 40(6):1913--1926, 2011.

\bibitem{spielman2011spectral}
D.~Spielman and S.~Teng.
\newblock Spectral sparsification of graphs.
\newblock {\em SIAM Journal on Computing}, 40(4):981--1025, 2011.

\bibitem{spielman2014sdd}
D.~Spielman and S.~Teng.
\newblock Nearly linear time algorithms for preconditioning and solving
  symmetric, diagonally dominant linear systems.
\newblock {\em SIAM Journal on Matrix Analysis and Applications},
  35(3):835--885, 2014.

\bibitem{steinberger2017globally}
M.~Steinberger, R.~Zayer, and H.-P. Seidel.
\newblock Globally homogeneous, locally adaptive sparse matrix-vector
  multiplication on the gpu.
\newblock In {\em Proceedings of the International Conference on
  Supercomputing}, page~13. ACM, 2017.

\bibitem{teng2016scalable}
S.-H. Teng.
\newblock Scalable algorithms for data and network analysis.
\newblock {\em Foundations and Trends{\textregistered} in Theoretical Computer
  Science}, 12(1--2):1--274, 2016.

\bibitem{xueqian:tcad15}
X.~Zhao, L.~Han, and Z.~Feng.
\newblock {A Performance-Guided Graph Sparsification Approach to Scalable and
  Robust SPICE-Accurate Integrated Circuit Simulations}.
\newblock {\em Computer-Aided Design of Integrated Circuits and Systems, IEEE
  Transactions on}, 34(10):1639--1651, 2015.

\bibitem{zhiqiang:dac17}
Z.~Zhao and Z.~Feng.
\newblock A spectral graph sparsification approach to scalable vectorless power
  grid integrity verification.
\newblock In {\em Proceedings of the 54th Annual Design Automation Conference
  2017}, page~68. ACM, 2017.

\bibitem{zhao:dac19}
Z.~Zhao and Z.~Feng.
\newblock Effective-resistance preserving spectral reduction of graphs.
\newblock In {\em Proceedings of the 56th Annual Design Automation Conference
  (DAC) 2019}, page 109. ACM, 2019.

\bibitem{zhiqiang:iccad17}
Z.~Zhao, Y.~Wang, and Z.~Feng.
\newblock {SAMG: Sparsified Graph Theoretic Algebraic Multigrid for Solving
  Large Symmetric Diagonally Dominant (SDD) Matrices}.
\newblock In {\em Proceedings of the 36th International Conference on
  Computer-Aided Design (ICCAD)}. ACM, 2017.

\bibitem{zhao2018nearly}
Z.~Zhao, Y.~Wang, and Z.~Feng.
\newblock Nearly-linear time spectral graph reduction for scalable graph
  partitioning and data visualization.
\newblock {\em arXiv e-print, arXiv:1812.08942}, 2018.

\end{thebibliography}
%\scriptsize
}

\end{document}